\documentclass{icrc}

\usepackage{graphicx} 
\newcommand{\psim}{\lower.5ex\hbox{$\; \buildrel \propto \over\sim \;$}}
\newcommand{\lesssim}{\lower.5ex\hbox{$\; \buildrel < \over\sim \;$}}
\newcommand{\gtrsim}{\lower.5ex\hbox{$\; \buildrel > \over\sim \;$}}

\newcommand{\e}{\epsilon}

\newcommand{\el}{\ell_{\rm S}}

\newcommand{\ep}{\epsilon^\prime}
\newcommand{\Ep}{E^\prime}
\newcommand{\apj}{Astrophys.~J.}
\newcommand{\apjl}{Astrophys.~J.}

\begin{document}

\title{Gamma Ray Bursts and Cosmic Ray Origin}
\author[1]{C. D. Dermer}
\affil[1]{Code 7653, Naval Research Laboratory, Washington, DC 20375-5352 USA}

\correspondence{dermer@gamma.nrl.navy.mil}

\firstpage{1}
\pubyear{2001}


\maketitle

\begin{abstract}
This paper presents the theoretical basis of the fireball/blast wave
model, and some implications of recent results on GRB source models
and cosmic-ray production from GRBs.  BATSE observations of the prompt
$\gamma$-ray luminous phase, and Beppo-SAX and long wavelength
afterglow observations of GRBs are briefly summarized. Derivation of
spectral and temporal indices of an adiabatic blast wave decelerating
in a uniform surrounding medium in the limiting case of a
nonrelativistic reverse shock, both for spherical and collimated
outflows, is presented as an example of the general theory. External
shock model fits for the afterglow lead to the conclusion that GRB
outflows are jetted. The external shock model also explains the
temporal duration distribution and clustering of peak energies in
prompt spectra of long-duration GRBs, from which the redshift
dependence of the GRB source rate density can be derived.  Source
models are reviewed in light of the constant energy reservoir result
of Frail et al.\ that implies a total GRB energy of a few $\times
10^{51}$ ergs and an average beaming fraction of $\approx 1/500$ of full sky.
Paczy\'nski's isotropic hypernova model is ruled out. The
Vietri-Stella model two-step collapse process is preferred over a
hypernova/collapsar model in view of the X-ray observations of GRBs
and the constant energy reservoir result. Second-order processes in
GRB blast waves can accelerate particles to ultra-high energies. GRBs
may be the sources of UHECRs and cosmic rays with energies above the
knee of the cosmic ray spectrum. High-energy neutrino and $\gamma$-ray
observations with GLAST and ground-based $\gamma$-ray telescopes will
be crucial to test GRB source models.
\end{abstract}

\section{Introduction}

In only ten years, the study of gamma-ray bursts has grown from a
narrow specialty involving a small group of high-energy astronomers to
a mainstream field of research that impacts nearly all topics in
astronomy. This change resulted primarily from observational advances
made with the Burst and Transient Source Experiment (BATSE) on the
{\it Compton Gamma Ray Observatory}, and the discovery of X-ray
afterglows and long wavelength counterparts made possible with the
Beppo-SAX experiment. The BATSE results showed that the angular
distribution of GRBs on the sky is isotropic, and that the GRB size
distribution exhibits a strong flattening for faint bursts
\citep{mee92}. This behavior follows from a cosmological origin of GRB
sources, with the decline in the number of faint bursts due to cosmic
expansion. Follow-up X-ray observations with the Narrow Field
Instruments on Beppo-SAX has permitted redshift determinations that
firmly establish the distance scale to the sources of $\gtrsim 2$ s
duration GRBs, which are those to which Beppo-SAX is sensitive.

The redshifts of nearly 20 GRBs are now known, with the mean redshift
near unity and the largest measured redshift at $z = 4.5$. The
corresponding distances imply apparent isotropic $\gamma$-ray energy
releases in the range from $\approx 10^{51}$-$10^{54}$ ergs. Recent
results suggest that GRB emissions are strongly beamed, so that the
total energy release is actually in the neighborhood of $10^{51}$
ergs, corresponding to the typical total energies released in the
kinetic outflow of a supernova. Delayed reddened enhancements detected
in the optical light curves of a few GRBs could also be a consequence
of a supernova emission component. If the beaming results are correct,
then many more sources of GRBs exist than are implied through direct
statistical studies of detected GRBs. The implied rate of both aligned
and misdirected GRB sources begins to approach the expected rate of
Type Ib/c supernovae (SNe). These and other lines of evidence indicate
that GRBs are related to a type of supernova.

This review of the theory of GRBs briefly touches on some of the
central questions in contemporary GRB studies. After summarizing the
prompt $\gamma$-ray luminous and delayed afterglow observations of
GRBs, detections of X-ray emission line and absorption features, and
some related phenomenology, the redshift distribution is
presented. These redshifts are obtained from absorption lines in the
optical afterglow, spectroscopy of the host galaxies and, in a few
cases, from X-ray line observations. Given the redshift, the directed
energy releases per unit solid angle can be derived. The theoretical
basis for the fireball/blast wave model is outlined, and numerical
calculations derived from the model are presented. According to the
blast-wave model, a central source ejects an expanding relativistic
shell of particles. Accelerated electrons radiate synchrotron photons
to produce the long wavelength X-ray, optical, and radio afterglows,
as well as the soft $\gamma$-ray emission emitted in the prompt phase
(although arguments have also been made for a Compton origin of this
emission). By contrast, a synchrotron self-Compton origin may explain
the $> 100$ MeV emission detected with EGRET from a handful of GRBs in
the prompt phase, and from GRB 940217 in the afterglow phase.

The photons received during the prompt phase are emitted closest to
 the central source; therefore a correct theory of the origin of the
 prompt emission reveals properties of the central engine.  We
 contrast an internal shock/colliding shell model with an external
 shock model for the prompt $\gamma$-ray emission. Beaming breaks in
 afterglow light curves can be used to deduce the collimation angle,
 and hence the true energetics of the GRB source. The most popular
 current model for the central sources of long-duration GRBs is the
 hypernova/collapsar model, where an evolved, massive star's central
 core collapses to a black hole while accreting material at the rate
 of $\approx 0.1$- a few Solar masses per second. The coalescence of
 neutron stars and black holes has been less successful as a model for
 the long-duration GRBs because host galaxy evidence indicate that a
 massive star origin is more likely; a compact-object coalescence
 model may still apply to the short duration GRBs. The supranova model
 of Vietri and Stella invokes a two-step collapse process where the
 core collapse of a massive star forms a rotationally stabilized
 neutron star following a supernova explosion. The neutron star later
 collapses to form a black hole. This model has several advantages
 over the hypernova/collapsar model to account for the evidence of
 beaming and the X-ray absorption and line observations, including the
 presence of a highly enriched circumburster medium (CBM), and a
 uniform energy scale set by the mass of the neutron star.

Some important issues in GRB cosmology are mentioned, though the
remainder of this review focuses on the role that GRBs may play in
cosmic-ray origin studies. GRB sources release energy into an $L^*$
galaxy such as the Milky Way at the time averaged rate of $\sim
10^{40\pm 1}$ ergs s$^{-1}$, which is nearly equal to the cosmic ray
power into our Galaxy. Relativistic blast waves can accelerate
particles well beyond the knee of the cosmic-ray spectrum through
second-order processes, overcoming difficulties that first-order Fermi
acceleration faces to accelerate cosmic rays to energies above the
knee of the cosmic ray spectrum in SNe. If the beaming results are
correct, then GRBs occur in our Galaxy at the rate of once every
several thousand years, suggesting that GRBs are associated with a
subset of SNe.  Finally, we consider high-energy neutrino production
from GRBs, and show how high-energy neutrino and gamma-ray
observations of GRBs can discriminate between an internal and external
shock model for the prompt gamma-ray emission, with important
implications for central engine theory.

\section{Observations and Phenomenology}

Because the emphasis here is on theory, we can only briefly summarize
the basic features of GRB observations and phenomenology (for a recent
review of the observations, see \cite{pkw00}). GRBs are brief flashes
of radiation at hard X-ray and soft $\gamma$-ray energies that display
a wide variety of time histories, though in $\sim$ 25\% of the cases a
characteristic single-pulse profile is observed, consisting of a rapid
rise followed by a slower decay \citep{fm95}. GRBs were first detected
at soft $\gamma$-ray energies with wide field-of-view
instruments. Peak soft $\gamma$-ray fluxes reach hundreds of photons
cm$^{-2}$ s$^{-1}$ in rare cases. The BATSE instrument is sensitive in
the 50-300 keV band, and provides the most extensive data base of GRB
observations during the prompt phase. It searches for GRBs by
examining strings of data for $> 5.5\sigma$ enhancements above
background on the 64 ms, 256 ms, and 1024 ms time scales, and triggers
on GRBs as faint as $\approx 0.5$ ph cm$^{-2}$ s$^{-1}$, corresponding
to energy flux sensitivities $\lesssim 10^{-7}$ ergs cm$^{-2}$
s$^{-1}$.

\begin{figure}[t]
\vskip-0.4in
\vspace*{10.0mm} 
\includegraphics[width=8.3cm]{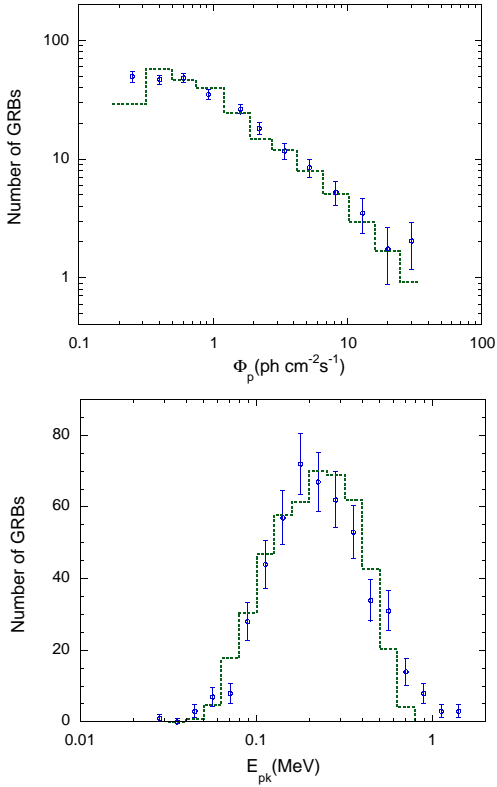} 
\caption{Differential distribution of the number of GRBs with peak fluxes 
measured on the 1024 ms time scale with BATSE is plotted in the upper
panel, and the distribution of the photon energies of the $\nu F_\nu$
peaks are plotted on the bottom panel. Histogram gives a fit to the
data using an external-shock model for the prompt phase.}
\label{fig0a}
\end{figure}

The integral size distribution of BATSE GRBs in terms of peak flux
$\phi_p$ is very flat below $\sim 3$ ph cm$^{-2}$ s$^{-1}$, and
becomes steeper than the $-3/2$ behavior expected from a Euclidean
distribution of sources at $\phi_p \gtrsim 10$ ph cm$^{-2}$ s$^{-1}$
\citep{pkw00}.  GRBs typically show a very hard spectrum in the hard
X-ray to soft $\gamma$-ray regime, with a photon index breaking from
$\approx -1$ at photon energies $E_{ph}\lesssim 50$ keV to a $-2$ to
$-3$ spectrum at $E_{ph} \gtrsim$ several hundred keV
\citep{ban93}. Consequently, the distribution of the peak photon
energies $E_{pk}$ of the time-averaged $\nu F_\nu$ spectra of BATSE
GRBs are typically found in the 100 keV - several MeV range
\citep{mal95}.  The differential 
peak-flux size distribution \citep{mee96} and the $E_{pk}$
distribution \citep{mal97} are shown by the data points in the top and
bottom panels of Fig.\ 1, respectively.

\begin{figure}[t]
\vskip-0.4in
\vspace*{10.0mm} 
\includegraphics[width=8.3cm]{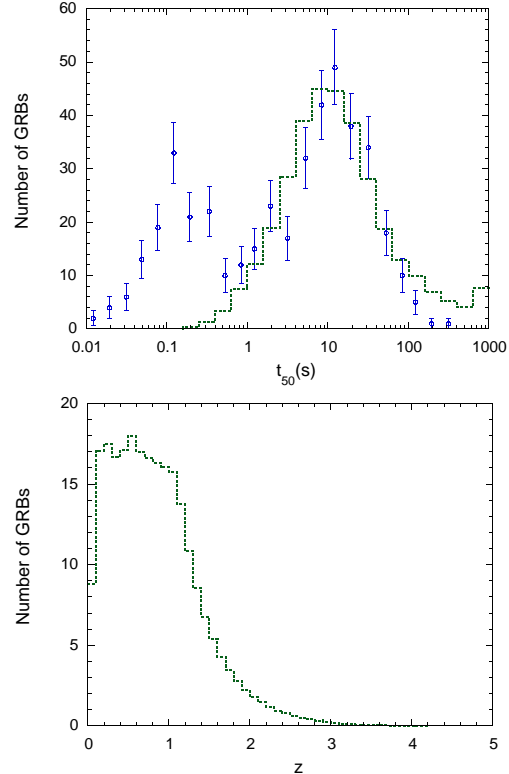} 
\caption{Data points in the top panel give the $t_{50}$ duration
distribution. External shock-model fit to the long duration
distribution is given by the histogram. Bottom panel shows the
redshift distribution of BATSE GRBs predicted by the model
\citep{bd00}.}
\label{fig0b}
\end{figure}

The duration of a GRB is defined by the time during which the middle
50\% ($t_{50}$) or 90\% ($t_{90}$) of the counts above background are
measured. A bimodal duration distribution is measured, irrespective of
whether the $t_{50}$ or $t_{90}$ durations are considered
\citep{kou93}. About two-thirds of BATSE GRBs are long-duration GRBs
with $t_{90}\gtrsim 2$ s, with the remainder comprising the
short-duration GRBs. The top panel of Fig.\ \ref{fig0b} shows the
$t_{50}$ duration distribution of BATSE GRBs.

The Beppo-SAX GRB observations revealed that essentially all
long-duration GRBs have fading X-ray afterglows \citep{cos99}. The
Wide Field Camera on Beppo-SAX has sensitivity down to $\sim 10^{-10}$
ergs cm$^{-2}$ s$^{-1}$ with $\lesssim 10^\prime$ error
boxes. Spacecraft slewing requires 6-8 hours, but permits Narrow Field
Instrument X-ray observations with sensitivity down to $\sim 10^{-14}$
ergs cm$^{-2}$ s$^{-1}$ and error boxes $\lesssim 0.5^\prime$.  The
first X-ray afterglow was obtained from GRB 970228 \citep{cos97},
which revealed an X-ray source which decayed according to a power-law
behavior $\phi_X\propto t^{\chi}$, with $\chi \sim -1.33$. Typically,
$\chi \sim -1.1$ to $-1.5$ in X-ray afterglow spectra.

\begin{figure}[t]
\vskip-1.8in
\vspace*{10.0mm} 
\includegraphics[width=8.3cm]{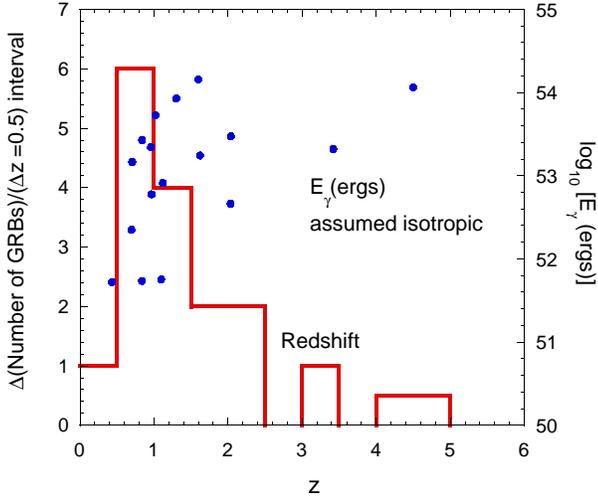} 
\caption{Distribution of redshifts and apparent isotropic $\gamma$-ray 
energy releases.}
\label{zandE}
\end{figure}

The small X-ray error boxes allow deep optical and radio follow-up
studies.  GRB 970228 was the first GRB from which an optical
counterpart was observed \citep{van97}, and GRB 970508 was the first
GRB for which a redshift was measured \citep{djo97,met97}.  Detection
of optical emission lines from the host galaxy, and absorption lines
in the fading optical afterglow due to the presence of intervening gas
has provided redshifts for about $20$ GRBs \citep{djo01}.  No optical
counterparts are detected from approximately one-half of GRBs with
well-localized X-ray afterglows, and are termed ``dark" bursts. These
sources may be undetected in the optical band because of dusty media
\citep{gro98}. Approximately 40\% of GRBs have radio counterparts, and
the transition from a scintillating to smooth behavior in the radio
afterglow of GRB 980425 provides evidence for an expanding source
\citep{fra97}. Fig.\ \ref{zandE} gives the distribution of redshift
and apparent isotropic $\gamma$-ray energy releases for 17 GRB for
which the redshift is known, using the compilation of \citet{fra01}.

X-ray emission lines, possibly due to Fe K$\alpha$ fluorescence, were
detected during a rebrightening phase in the afterglow 
of GRB 970508 \citep{pir99}, and
in the afterglow spectra of GRB 991216 \citep{pir00} and GRB 000214
\citep{ant00}.  A marginal detection of an X-ray emission feature was
also reported from GRB 970828 \citep{yos99}.  A transient absorption
feature was detected from GRB 990705 \citep{ama00}, and X-ray
absorption in excess of the Galactic hydrogen column density has also been
reported in GRB 980329 \citep{fro00}. These results provide an
alternative method of redshift determination, and provide important
clues about the environment of the progenitor object, showing that
large quantities of iron must be present in the vicinity of these
sources.

The EGRET instrument on the {\it Compton Observatory} detected 7 GRBs
with $> 30$ MeV emission during the prompt phase \citep{cds98},
including the extraordinary burst GRB 940217, which displayed $\gtrsim
100$ MeV emission 90 minutes after the onset of the GRB, including one
photon with energy near 20 GeV \citep{hur94}. This gives unambiguous
evidence for the importance of nonthermal processes in GRBs. TeV
radiation has been reported to be detected with the Milagrito water
Cherenkov detector from GRB 970417a \citep{atk00}. If correct, this
requires that this source be located at $z \lesssim 0.3$ in order that
$\gamma\gamma$ attenuation with the diffuse intergalactic infrared
radiation to be small.

A class of X-ray rich GRBs, with durations of order of minutes and
X-ray fluxes in the range $10^{-8}$-$10^{-7}$ ergs cm$^{-2}$ s$^{-1}$
in the 2-25 keV band, has been detected with many X-ray satellites,
including Ariel V, HEAO-1, ROSAT, Ginga, and Beppo-SAX
\citep{hei01}. We also note several phenomenological correlations,
including claims of an inverse correlation of luminosity and BATSE
channel lags \citep{nor00}, correlations between the duration of
quiescent and subsequent pulse periods in separated pulses
\citep{ram01}, and indications that the peak luminosity is correlated
with variability \citep{rei01}. Other phenomenological behaviors in
the prompt emission include hard-to-soft evolutions and
hardness-intensity correlations, and a quantitative relation between
integrated fluence and $E_{pk}$ in well-defined pulses \citep{lk96}.

\section{Elementary Blast Wave Theory}

The fireball/blast-wave model for GRBs was proposed by
\citet{rm92} and \citet{pr93}, and successfully predicted power-law
afterglows \citep{mr97}. It has emerged as the standard model to
interpret GRB prompt and afterglow emissions (for a recent review, see
\citet{mes02}).  The simplest version of the standard blast wave model
involves a spherical, uncollimated explosion taking place in a uniform
surrounding medium.  A relativistic pair fireball is formed when an
explosion deposits a large amount of energy into a compact volume. The
pressure of the explosion causes the fireball to expand, with the
thermal kinetic energy of the explosion being transformed into
directed kinetic energy due to strong adiabatic cooling of particles
in the comoving frame \citep{mlr93,piran99}.  Because of the Thomson
coupling between the particles and photons, most of the original
explosion energy is carried by the baryons that were originally mixed
into the explosion. Under certain conditions involving less energetic,
temporally extended, or very baryon-clean explosions, neutrons can
decouple from the flow \citep{dkk99}. If this does not occur, then the
coasting Lorentz factor $\Gamma_0 \cong E_0/M_b c^2$ , where $M_b$ is
the baryonic mass and $E_0= 10^{52}E_{52}$ ergs is the apparent
isotropic energy release.

In the simplest version of the model, the blast wave is approximated
by a uniform thin shell, and particle acceleration is assumed to take
place at the forward shock only. A forward shock is formed when the
expanding shell accelerates the external medium, and a reverse shock
is formed due to deceleration of the cold shell. The forward and
reverse shocked fluids are separated by a contact discontinuity and
have equal kinetic energy densities. From the relativistic shock jump
conditions \citep{sp95}, $4\Gamma(\Gamma-1)n_0 =
4\bar\Gamma(\bar\Gamma-1)n^\prime_{sh}$, where $\Gamma$ is the blast
wave Lorentz factor, $\bar \Gamma$ is the Lorentz factor of the
reverse shock in the rest frame of the shell, $n^\prime_{sh}$ is the
density of the unshocked fluid in the proper frame of the expanding
shell, and $n_0$ is the density of the CBM, here assumed to be
composed of hydrogen. Particle acceleration at the reverse shock is
unimportant when the reverse shock is nonrelativistic, which occurs
when
\begin{equation}
\Gamma \ll \sqrt{n^\prime_{sh}\over n_0
}\;.
\label{NRS}
\end{equation}

The unique feature of GRBs is that the coasting Lorentz factor
$\Gamma_0$ may reach values from hundreds to thousands. By contrast,
Type Ia and II SNe have ejecta speeds that are in the range of $\sim
5000$-$30000$ km s$^{-1}$. The relativistic motion of the radiating
particles introduce many interesting effects in GRB emissions that
must be properly taken into account.

Three frames of reference are considered when discussing the emission
from systems moving with relativistic speeds: the stationary frame,
which are denoted here by asterisks, the comoving frame, denoted by
primes, and the observer frame. The differential distance traveled by
the expanding blast wave during differential time $dt_*$ is simply $dx
= \beta c dt_*$, where $\beta = \sqrt{1-\Gamma^{-2}}$. Due to time
dilation, $dx = \beta \Gamma c dt^\prime$. Because of time dilation,
the Doppler effect, and the cosmological redshift $z$, the
relationship between comoving and observer times is $(1+z)\Gamma
dt^\prime (1-\beta\cos\theta) = (1+z)dt^\prime/\delta = dt$, where
$\theta$ is the angle between the emitting element and the observer
and $\delta = [\Gamma(1-\beta\cos\theta)]^{-1}$ is the Doppler
factor. For on-axis emission from a highly relativistic emitting
region, we therefore see that $dt \cong (1+z)dx /\Gamma^2 c$;
consequently the blast wave can travel a large distance $\Gamma^2 c
\Delta t$ during a small observing time interval. A photon measured
with dimensionless energy $\epsilon = h\nu/m_ec^2$ is emitted with
energy $\delta\epsilon^\prime/(1+z)$.

A blast wave expanding into the surrounding medium acts as a fluid if
there is a magnetic field in the comoving frame sufficiently strong to
confine the particles within the width of the blast wave.  The
blast-wave width depends on the duration of the explosion and the
spreading of the blast-wave particles. Because sub-second structure is
regularly observed in GRB light curves, the central engine must be
$\lesssim 1$ lt-s, and is probably much thinner in view of 1-10 ms
variability observed in some GRBs \citep{bha92}. We denote this length
scale by $\Delta_0$ cm. The shell will spread radially in the comoving
frame by an amount $x_\parallel^\prime \simeq v_{spr} t^\prime$, where
$t^\prime \cong x/(\beta\Gamma c)$ is the available comoving time and
$v_{spr}$ is the spreading speed. This implies a shell width in the
observer frame of $\Delta = x/\Gamma_0^2$ due to length contraction,
and a spreading radius $x_{spr} = \Gamma_0^2 \Delta_0$, assuming that
the shell spreads with speed $v_{spr} \cong c$. The width of the
unshocked blast-wave fluid in the rest frame of the explosion is
therefore $\Delta = \min(\Delta_0, x/\Gamma_0^2$)
\citep{mlr93,snp96}. Milligauss fields are sufficient to confine
relativistic electrons.

As the blast wave expands, it sweeps up material from the surrounding
medium to form an external shock \citep{mr93}. In a colliding wind
scenario, the blast wave intercepts other portions of the relativistic
wind \citep{rm94}. Protons captured by the expanding blast wave from
the CBM will have total energy $\Gamma m_p c^2$ in the fluid frame,
where $m_p$ is the proton mass. The kinetic energy swept into the
comoving frame by an uncollimated blast wave at the forward shock per
unit time is given by
\begin{equation}
{dE^\prime\over dt^\prime} = 4\pi x^2 n_0 m_pc^3
\beta\Gamma(\Gamma-1)
\label{dEdt}
\end{equation} 
\citep{bm76}.  One factor of $\Gamma$ represents the increase of
external medium density due to length contraction, the factor ($\Gamma
-1$) is proportional to the kinetic energy of the swept up particles,
and the factor $\beta$ is proportional to the rate at which the
particle energy is swept. Thus the power is $\propto \Gamma^2$ for
relativistic blast waves, and $\propto \beta^3$ for nonrelativistic
blast waves. This process provides internal energy available to be
dissipated in the blast wave.

A proton that is captured by and isotropized in the comoving frame
will have energy $\Gamma m_p c^2$ in the comoving fluid frame, or
energy $\Gamma^2 m_p c^2$ in the observer frame. The expanding shell
will therefore begin to decelerate when $E_0 = \Gamma_0 M_b c^2 =
\Gamma_0^2 m_p c^2 (4\pi x_d^3 n_0/3) $, giving the deceleration
radius
\begin{equation} 
x_d \equiv ( {3 E_0\over
4\pi\Gamma_0^2 c^2 m_p n_0})^{1/3} \cong 2.6\times 10^{16}
({E_{52}\over \Gamma_{300}^2 n_0})^{1/3}\;\rm{cm}\;
\label{x_d}
\end{equation}
\citep{rm92,mr93}, where $\Gamma_{300} = \Gamma_0/300$.
 Acceleration at the shock front can inject power-law distributions of
particles. In the process of isotropizing the captured particles,
 magnetic turbulence is introduced \citep{ps00} that can also accelerate
particles to very high energies through a second-order Fermi process.

The deceleration time as measured by an on-axis observer is given by
\begin{equation}
t_d \equiv (1+z) {x_d\over \beta_0 \Gamma_0^2 c}\;\cong {9.6\;
(1+z)\over \beta_0}\;({E_{52}\over \Gamma_{300}^8
n_0})^{1/3}\;\rm{s}\;.
\label{t_d}
\end{equation}
The factor $\beta_0^{-1} = 1/\sqrt{1-\Gamma_0^{-2}}$ generalizes
\citep{dh01} the result of \citet{mr93} for mildly relativistic and
nonrelativistic ejecta, as in the case of Type Ia and Type II
supernova explosions. The Sedov radius
\begin{equation}
\ell_{\rm S} = \Gamma_0^{2/3} x_d = 
( {3 E_0\over 4\pi m_p c^2  n_0})^{1/3} \cong \label{l_S}
\end{equation}
$$
1.2\times 10^{18} ({E_{52}\over n_0})^{1/3}\;{\rm~cm}\;
\cong 6.6\times 10^{18} ({m_\odot\over n_0})^{1/3}\;{\rm~cm}\;,$$
where $m_\odot$ is the total (rest mass plus kinetic) explosion energy
in units of Solar rest mass energy.  For relativistic ejecta,
$\ell_{\rm S}$ refers to the radius where the blast wave slows to
mildly relativistic speeds, i.e., $\Gamma
\sim 2$. The Sedov radius of a SN that ejects a 10 $M_\odot$ envelope
could reach $\sim 5$ pc or more. The Sedov age $t_{\rm S} = \ell_{\rm
S}/v_0 \cong 700 (m_\odot/n_0)^{1/3}/(v_0/0.01 c)$ yr for
nonrelativistic ejecta, and is equivalent to $t_d$ in general.

The evolution of an adiabatic blast wave in a
uniform surrounding medium for an explosion with
a nonrelativistic reverse shock is given by \citep{dh01}
\begin{equation}
P(x) = {P_0\over \sqrt{1+(x/x_d)^3}}\cong 
 \cases{\beta_0\Gamma_0 \; ,& $x\ll x_d$ \cr\cr
        \beta_0 ({x\over \el})^{-3/2}\; , & $x_d \ll x $ \cr}\;,
\label{P(x)}
\end{equation}
where $P = \beta\Gamma$ and $P_0 = \beta_0\Gamma_0$
 represent dimensionless bulk momenta of the shocked fluid. 
Equation (\ref{P(x)}) reduces to the adiabatic Sedov behavior for
nonrelativistic ($\beta_0\Gamma_0 \ll 1$) explosions, giving $v
\propto x^{-3/2}$, $x \propto t^{2/5}$, and $v\propto t^{-3/5}$, as is
well-known. In the relativistic ($\Gamma_0 \gg 1$) limit, $\Gamma
\propto x^{-3/2}$ and $t \cong c^{-1} \int dx/\Gamma^2 \propto \int dx
\; x^3$, yielding $x \propto t^{1/4}$ and $\Gamma \propto t^{-3/8}$
when $\Gamma \gg 1$. The adiabatic behavior does not apply when
radiative losses are important. A nonrelativistic supernova shock
becomes radiative at late stages of the blast-wave evolution. It is
not clear whether a GRB fireball is highly radiative or nearly
adiabatic during either its gamma-ray luminous prompt phase or during
the afterglow \citep{wax97,vie97}.  In the fully radiative limit,
equation of energy conservation gives $\beta_0\Gamma_0 M_0 \cong \beta
\Gamma[M_0 + 4\pi n_0 x^3/3 ]$, giving the asymptotes $\Gamma \propto
x^{-3}$ when $\Gamma_0 \gg \Gamma\gg 1$ and $\beta \propto x^{-3}$
when $\Gamma - 1 \ll 1$.

Most treatments employing blast-wave theory to explain the observed
radiations from GRBs assume that the radiating particles are
electrons. The problem here is that $\sim m_p/m_e \sim 2000$ of the
nonthermal particle energy swept into the blast-wave shock is in the
form of protons or ions unless the surroundings are composed primarily
of electron-positron pairs
\citep{kg01}. For a radiatively efficient system, physical
processes must therefore transfer a large fraction of the swept-up
energy to the electron component. In elementary treatments of the
blast-wave model, it is simply assumed that a fraction $e_e$ of the
forward-shock power is transferred to the electrons, so that
\begin{equation}
L_e^\prime = e_e {dE^\prime\over dt^\prime} \;.
\label{Le}
\end{equation}

If all the swept-up electrons are accelerated, then joint
normalization to power and number gives
\begin{equation}
\gamma_{min} \cong e_e({p-2\over p-1})({m_p\over
m_e})(\Gamma-1)\;\cong e_e k_p({m_p\over m_e})\Gamma
\label{gminap}
\end{equation}
for $2 < p < 3$, where the last expression holds when $\Gamma\gg 1$
and $k_p = (p-2)/(p-1)$.
 
The strength of the magnetic field is another major uncertainty. The
standard prescription is to assume that the magnetic field energy
density $u_b = B^2/8\pi$ is a fixed fraction $e_B$ of the downstream
energy density of the shocked fluid. Hence
\begin{equation}
{B^2\over 8\pi} = 4e_B n_0 m_pc^2(\Gamma^2-\Gamma)\;.
\label{B^2}
\end{equation}
It is also generally supposed in simple blast-wave model calculations
that some mechanism -- probably the first-order shock Fermi process --
injects electrons with a power-law distribution between electron
Lorentz factors $\gamma_{min} \leq \gamma \leq \gamma_2$ downstream of
the shock front.  The
electron injection spectrum in the comoving frame
is modeled by the expression
\begin{equation}
{dN_e^\prime(\gamma)\over dt^\prime d\gamma} = K_e
\gamma^{-p}\;,\; {\rm for~} \gamma_{min}< \gamma < \gamma_{2}\;,
\label{dNedgdt}
\end{equation}
where $K_e$ is normalized to the rate at which electrons are captured.

The maximum injection energy is obtained by balancing synchrotron
losses and an acceleration rate given in terms of the inverse of the
Larmor time scale through a parameter $e_{2}$, giving
\begin{equation}
\gamma_2 \cong 4\times 10^7 e_{2}/\sqrt{B({\rm G})}\;.
\label{gammamax}
\end{equation} 
A break is formed in the electron spectrum at cooling electron Lorentz
factor $\gamma_c$, which is found by balancing the synchrotron loss
time scale $t_{syn}^\prime$ with the adiabatic expansion time
$t^\prime_{adi} \cong x/\Gamma c\cong \Gamma t \cong t_{syn}^\prime
\cong (4c\sigma_{\rm T} B^2 \gamma_c/24\pi m_e c^2)^{-1}$, giving
\begin{equation}
\gamma_c \cong {3m_e\over 16 e_B n_0 m_p c\sigma_T\Gamma^3 t}\;
\label{gamma_c}
\end{equation}
\citep{spn98}. For an adiabatic blast wave, $\Gamma\propto t^{-3/8}$, so that 
$\gamma_{min} \propto t^{-3/8}$ and $\gamma_c \propto t^{1/8}$.

The observed $\nu F_\nu$ synchrotron spectrum from a GRB depends on
the geometry of the outflow. Denoting the comoving spectral luminosity
by $L^\prime_{syn}(\ep )$ $=$ $\ep (dN^\prime/d\ep dt^\prime)$, then
$\epsilon^\prime L^\prime _{syn}(\ep)
\cong {1\over 2} u_B c\sigma_{\rm T} \gamma^3 N_e(\gamma)$, 
with $\gamma = \sqrt{\ep/\e_B}$ and $\e_B = B/B_{cr} = B/(4.41\times
10^{13}$ G).  For a spherical blast-wave geometry, the spectral power
is amplified by two powers of the Doppler factor $\delta$ for the
transformed energy and time.  The $\nu F_\nu$ synchrotron spectrum is
therefore
\begin{equation}
f_\epsilon^{syn}\cong {2\Gamma^2\over 4\pi d_L^2}\;(u_Bc\sigma_{\rm
T})\; \gamma^3 N^\prime_e(\gamma)\;,\;\gamma \cong \sqrt{{(1+z)\e\over
2\Gamma\e_B}}\;.\;
\label{fe}
\end{equation}
where $d_L$ is the luminosity distance.

For collimated blast waves with jet opening angle $\theta_j$, equation
(\ref{fe}) does not apply when $\Gamma\lesssim 1/\theta_j$ because
portions of the blast wave's radiating surface no longer contribute to
the observed emission.  In this case, the blast-wave geometry is a
localized emission region, and the received $\nu F_\nu$ spectrum is
given by
\begin{equation}
f^{syn}_\epsilon\cong {\delta^4\over 4\pi d_L^2}\;({1\over
2}u_Bc\sigma_{\rm T})\; \gamma^3 N^\prime_e(\gamma)\;,\;\gamma \cong
\sqrt{{(1+z)\e\over \delta\e_B}}\;.\;
\label{feblob}
\end{equation}
Here the observed flux is proportional to four powers of the Doppler
factor: two associated with solid angle, one with energy and one with
time.

From this formalism, analytic and numerical models of relativistic
blast waves can be constructed. It is useful to define two regimes
depending on whether $\gamma_{min} < \gamma_c$, which is called the
weak cooling regime, or $\gamma_{min} > \gamma_c$, called the strong
cooling regime \citep{spn98}. If the parameters $p$, $e_e$ and $e_B$
remain constant during the evolution of the blast wave, then a system
originally in the weak cooling regime will always remain in the weak
cooling regime. In contrast, a system in the strong cooling regime
will evolve to the weak cooling regime.

For a power-law injection spectrum given by equation (\ref{dNedgdt}),
the cooling comoving nonthermal electron spectrum can be approximated
by
\begin{equation}
N^\prime_e(\gamma) \cong  {N_e^0\;\gamma_0^{s-1}\over s-1}
\cases{\gamma^{-s}\; ,&  $\gamma_0 \lesssim \gamma \lesssim \gamma_1$ \cr\cr
	\gamma_1^{p+1-s}\gamma^{-(p+1)} \; , & $\gamma_1 \lesssim
\gamma \lesssim \gamma_2,$ \cr}
\label{Ne}
\end{equation} 
and $N_e^0=4\pi x^3 n_0/3$ for the assumed system.
In the weak cooling regime, $s = p$, $\gamma_0 = \gamma_{min}$ and
$\gamma_1 = \gamma_c$, whereas in the strong cooling regime, $s = 2$,
$\gamma_0 = \gamma_c$ and $\gamma_1 = \gamma_{min}$.

\begin{figure}[t]
\vskip-0.5in
\vspace*{8.0mm} 
\includegraphics[width=8.3cm]{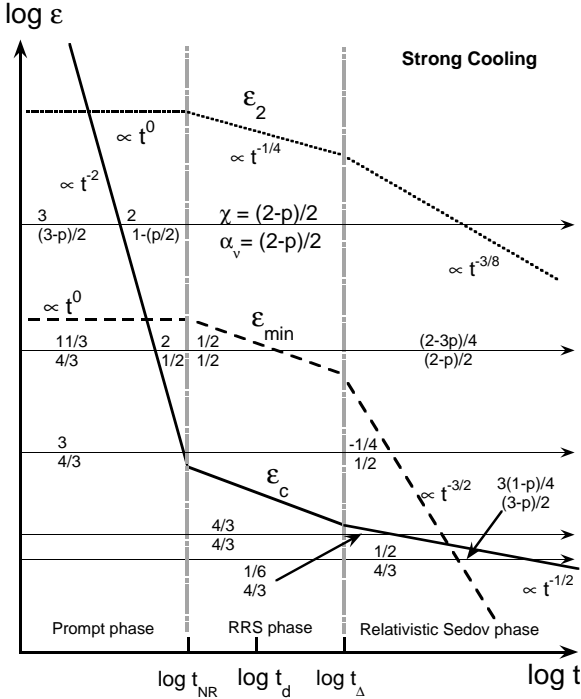} 
\vskip-0.3in
\caption{Possible time profiles and spectral behavior of the
nonthermal synchrotron radiation that could arise for a system that is
in the fast cooling regime during a portion of the adiabatic evolution
of a blast wave in a uniform surrounding medium. The temporal indices
$\chi$ and $\nu F_\nu$ spectral indices $\alpha_\nu$ are shown,
respectively, above and below the lines that represent different
families of possible emission trajectories.}
\label{anal}
\end{figure}

As an example of the elementary theory, consider the temporal index in
the strongly cooled regime for high energy electrons with $\gamma\geq
\gamma_1=\gamma_m\propto\Gamma$, $\gamma_0 = \gamma_c\propto
1/(\Gamma^3 t)$, and $s=2$. From equation (\ref{fe}), $f_\e \propto $
$\Gamma^4 x^3 $
$\gamma_0^{s-1}\gamma_1^{p+1-s}\e^{(2-p)/2}$ $/\Gamma^{2-p}$ $\propto
\Gamma^{2(p-1)}x^3\e^{(2-p)/2}/t$ $\propto$ $ t^\chi\e^\alpha_\nu$,
with $\chi = (2-3p)/4$ and $\alpha_\nu = (2-p)/2$. A few other such
results include the decay of a strong and weak cooling $\nu F_\nu$
peak frequency $\e_{pk}\propto t^{-3/2}$ and $\e_{pk}
\propto t^{-3p/2}$, resepctively. A cooling index change
 by $\Delta(\alpha_\nu) = 1/2$ 
is accompanied by a change of temporal index from $\chi =3(1-p)/4$ to
$\chi = (2-3p)/4$ at late times, so that $\Delta\chi = 1/4$. Extension
to power-law radial electron profiles is obvious, and beaming breaks
introduce two additional factors of $\Gamma$ from $\delta \approx
\Gamma$ in equation (\ref{feblob}) when the observer sees beyond the
causally connected regions of the jetted blob, noting that $N_e^0$
must be renomalized appropriately.

Fig.\ \ref{anal} shows spectral and temporal indices that are derived
from the preceding analytic considerations of a spherical adiabatic
blast wave that is in the strong cooling regime during a portion of
its evolution \citep{db02}. We use the notation that the $\nu F_\nu$
spectrum is described by
\begin{equation}
f_\e \propto  t^\chi \e^{\alpha_\nu}\;.
\label{f_e}
\end{equation}
Observations at a specific photon energy $\e$ will detect the system
evolving through regimes with different spectral and temporal
behaviors. The relativistic Sedov phase corresponds to the afterglow
regime, and we also consider in this figure a possible relativistic
reverse shock (RRS) phase. If equation (\ref{NRS}) is satisfied with
$\Gamma$ replaced by $\Gamma_0$, then the blast wave will not evolve
through the RRS phase. Even for this simple system, a wide variety of
behaviors are possible. Inclusion of additional effects, including a
nonuniform external medium \citep{mrw98} or blast-wave evolution in a
partially radiative phase \citep{bd00a} will introduce additional
complications.

The heavy solid and dotted lines in Fig.\ \ref{anal} represent the
evolution of the cooling photon energies $\e_c$ and $\e_{min}$,
respectively. In the lower-right hand corner of the diagram, one sees
that observations in the afterglow phase may detect a transition from
the uncooled portion of the synchrotron spectrum where $\chi =
(3-p)/2$ and $\alpha_\nu = (3-p)/2$ to a cooled portion of the
synchrotron spectrum where $\chi = (2-3p)/4$ and $\alpha_\nu =
(2-p)/2$. Thus a change in photon index by 0.5 units is accompanied by
a change in temporal index by 1/4 unit, independent of $p$.
Comparison with afterglow observations can be used to test the
blast-wave model \citep{gal98}. The upcoming Swift telescope will be
able to monitor behaviors in the time interval between the prompt and
afterglow phase, and to search for other regimes of blast wave
evolution and evidence for evolution in the non-adiabatic regime.

\begin{figure}[t]
\vskip-0.5in
\vspace*{10.0mm} 
\includegraphics[width=8.3cm]{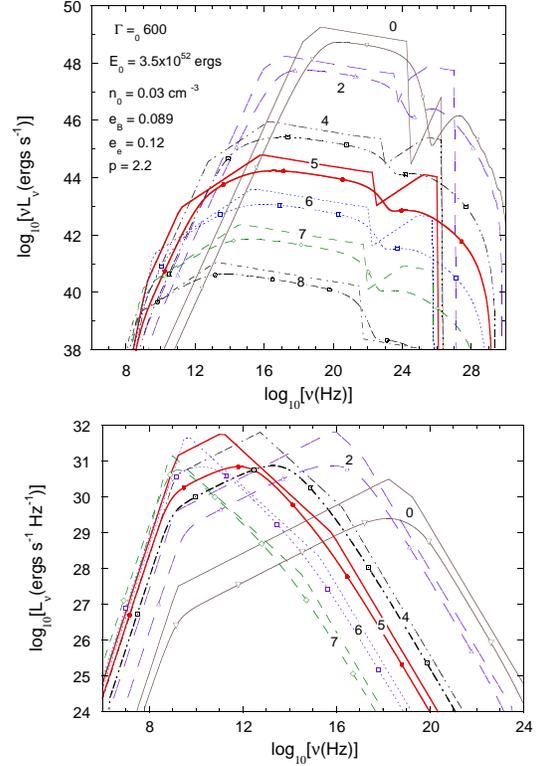} 
\vskip-0.2in
\caption{Calculations of spectral energy distributions emitted by a
relativistic blast wave with initial Lorentz factor $\Gamma_0 = 600$
that is energized by sweeping up material from a uniform surrounding
medium with density $n$. Curves are labeled by the base 10 logarithm
of the observing time in seconds.  The analytic and numerical models
are shown, with the latter curves identified by data points. Other
parameters of the calculation are shown in the legend, based on
parameters taken from the fit to GRB 970508 by \citet{wg99}. Upper
panel shows spectral energy distribution in a $\nu L_\nu$
representation, and the lower panel shows light curves at different
energies in an $L_\nu$ representation.}
\label{wg}
\end{figure}

Fig.\ \ref{wg} shows a numerical model of a relativistic blast wave
calculated with the code developed by \citet{cd99}, which takes into
account light travel-time delays from different portions of the blast
wave, synchrotron and synchrotron self-Compton emission, as well as
self-absorption. The numerical results are compared with analytic
spectral energy distributions \citep{dbc00} in which an SSC component
is included in the approach of \citet{spn98}. We use parameters
derived by \citet{wg99} to fit afterglow data from GRB 970508. Here
the total energy injected into the fireball is $E= 3.5\times 10^{52}$
ergs, and the density of the external medium is $n = 0.03$ cm$^{-3}$.
The SSC component becomes increasingly important with increasing
$e_e/e_B$ \citep{snp96,msb00}, and will be observed as features in the
X-ray and optical light curves \citep{dcm00a,zm01}. As can be seen,
the analytic expressions provide a reasonable representation of the
numerical model in the optical/X-ray regime, but the discrepancies at
radio frequencies are significant, as is clear from the second panel
in Fig.\ \ref{wg}. Although the times of the temporal breaks arising
from synchrotron emission emitted by the lowest energy electrons are
in good agreement with the numerical calculation, the analytic model
can be discrepant in flux values by an order-of-magnitude or more, so
that any quantitative statement based upon analytic models must be
made with care, particularly in the radio regime.
 
Fig.\ 6 shows light curves calculated from an adiabatic blast wave
that expands into a uniform surrounding medium \citep{dbc99}. The
light curves display a temporal profile that rapidly rises and slowly
decays, and reproduces the so-called fast-rise, exponential decay
profiles observed in a significant fraction of GRBs. This indicates
that the prompt emission in these GRBs, at least, arises from an
external shock model. An external shock origin also accounts for the
basic phenomenology of the GRBs with smooth profiles, including the
hard-to-soft evolution and the hardness-intensity correlation observed
in GRB light curves.

\begin{figure}[t]
\vskip-1.8in
\vspace*{0.0mm} 
\hspace*{-5.0mm}
\includegraphics[width=9.0cm]{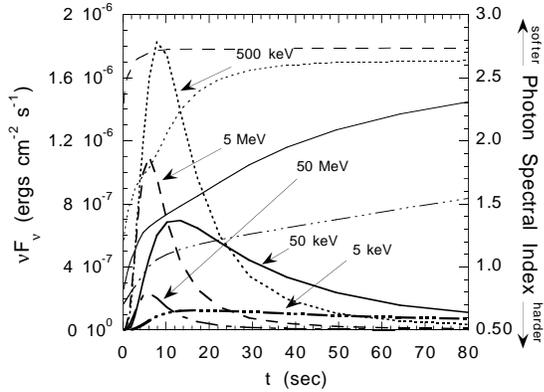} 
\vskip-0.7in
\caption{Numerical calculations of light curves measured at different
observed photon energies. These profiles correspond to the most common
prompt GRB light curves. The hard-to-soft evolution is indicated by
the variation in the photon index. For this model, an explosion with
total energy $E_0 = 10^{54}$ ergs and coasting Lorentz factor
$\Gamma_0 = 300$ expands into a uniform surrounding medium with
density $n_0 = 100$ cm$^{2}$. }
\label{fred}
\end{figure}

\subsection{Clean and Dirty Fireballs}

If a subset of prompt GRB light curves originate from an external
shock model in the prompt phase, one might ask what this means for the
remainder of the GRBs. A majority of the GRB community believes that
an active central engine is required to reproduce the spiky, short
timescale variablity observed in a large number of GRB light curves,
and that an impulsive explosion cannot produce the temporal gaps
detected in some GRB light curves. Inhomogeneities in the surrounding
medium can, in fact, produce short timescale variability and gaps in
the light curves, as demonstrated by
\citet{dm99}. More serious issues for the external shock model for the prompt 
emission are posed by the observations of optical emission observed in
GRB 990123 in the prompt phase if due to reverse shock emission
\citep{ake99}, and the correlated durations of quiescent and
subsequent peak durations in GRB light curves \citep{ram01}.

Rather than revisit this controversy, it seems more useful to focus on
the predictive capability of the external shock model. Because of the
small number of parameters, it is simple to outline the generic
behavior of GRBs in terms of the model's most sensitive parameters,
namely the total explosion energy $E_0$, the surrounding medium
density $n_0$, and the coasting Lorentz factor $\Gamma_0$.

The basic features of GRB fireballs can be summarized by a few simple
equations \citep{dcb99}.  The first is that the total energy liberated
in photons during the prompt phase is
\begin{equation}
E_\gamma \cong 0.1 \;\Pi_0 t_d\;,
\label{Egamma}
\end{equation}
where the precise value of the coefficient, here taken to be 10\%,
depends on how radiative the fireball is during this phase.  The power
$\Pi_0$ during the prompt phase is
\begin{equation}
\Pi_0 \propto (n_0 E_0^2)^{1/3}\; \Gamma_0^{8/3}\;
\label{Pi0}
\end{equation}
and, as already seen in equation (\ref{t_d}), the deceleration time 
\begin{equation}
t_d \propto ({E_0\over n_0})^{1/3}\;\Gamma_0^{-8/3}\;.
\label{td1}
\end{equation}
Most of the emission during the prompt phase is radiated at photon energy
\begin{equation}
E_{pk}\propto  n_0^{1/2}\Gamma_0^4\;.
\label{Epk}
\end{equation}

\begin{figure}[t]
\vskip-0.5in
\vspace*{0.0mm} 
\includegraphics[width=8.3cm]{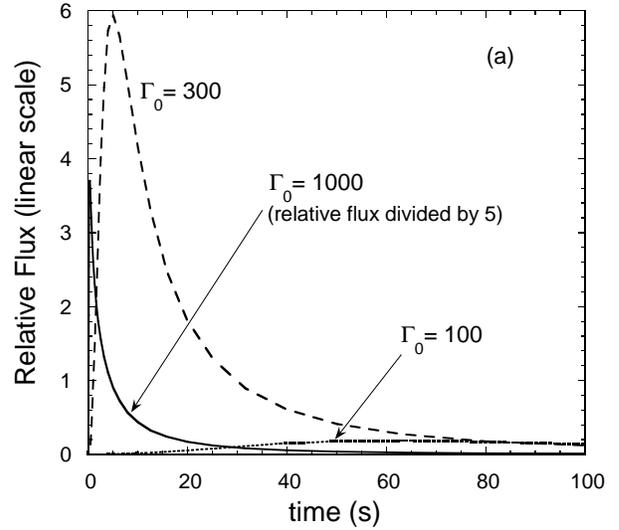} 
\vskip-1.0in
\caption{Light curves calculated at 100 keV from a fireball with
other parameters the same as in Fig.\ \ref{fred}. }
\label{Gammadep}
\end{figure}

The strong dependence of these quantities on $\Gamma_0$ is evident.  A
detector observing in a fixed range of photon energy centered at
$E_{det}$, such as BATSE, will have strong selection biases to detect
fireballs with different $\Gamma_0$ factors. The biases against
detecting dirty fireballs, namely those with small values of
$\Gamma_0$ and large baryon-loading, arise because the peak luminosity
measured by a detector declines rapidly when $\Gamma_0 \lesssim \bar
\Gamma_0$. The quantity $\bar\Gamma_0$ is the Lorentz factor of a
fireball that produces the bulk of its emission near the sensitive
range of a detector, and is defined by $E_{pk} = E_{det} \propto
n_0^{1/2}\bar\Gamma_0^4$, where $E_{det}$ is the photon energy to
which the GRB detector is largest effective area. The peak luminosity
is larger for clean fireballs with $\Gamma_0 > \bar \Gamma_0$ and
small baryon loading, but the shorter durations of the clean events
mean that detectors trigger on fluence rather than peak flux when
$\Gamma_0 \gg \bar \Gamma_0$. GRB detectors are therefore also biased
against the detection of clean fireballs. This explains the tendency
of detectors to detect GRBs that produce prompt radiation with $\nu
F_\nu$ peak in the waveband at which the detector has greatest
sensitivity.

These considerations resolve the apparent paradox that a beamed
scenario should seem to produce a wide range of $E_{pk}$, contrary to
observations \citep{dbc99} (see lower panel in Fig.\ \ref{fig0a}). To
avoid fine-tuning of $\Gamma_0$, one realizes that there must be two
additional classes of explosive phenomena that remain to be discovered
by detectors with appropriate design \citep{dcb99}, namely the large
baryon-load, low $\Gamma_0$ ``dirty" fireballs, and the low
baryon-load, high $\Gamma_0$ ``clean" fireballs.

This is apparent from Fig.\ \ref{Gammadep}, which shows how a
variation of a factor of 3 in $\Gamma_0$ from 300 to 100 can make a
GRB essentially undetectable in the BATSE band. The low Lorentz
factor, dirty fireballs are more extended in time, and radiate their
energy at lower observed photon frequencies. These may correspond to
the X-ray rich GRBs that have been detected with Beppo-SAX and
previous X-ray detectors \citep{hei01}.

By taking the triggering properties of BATSE into account, the
external shock model can be used to fit the statistical data on GRBs
\citep{bd00}.  The results are shown in Figs.\ \ref{fig0a} and
\ref{fig0b}, where simple power-law distributions of $\Gamma_0$ and
$E_0$ were assumed. The underlying assumption of this model is that
the rate of GRBs per comoving density follows the star formation
history of the universe. These results show that fine-tuning is not
required to explain the tendency of $E_{pk}$ to be within a narrow
range. The tendency of softer bursts to have longer durations is a
direct consequence of this model, though only the long duration GRBs
are fit in this model. A wide range of apparent isotropic energy
releases $E_0$ is used to fit the statistical data. If GRBs are
beamed, the distribution of $E_0$ would be consistent with recent
analyses that indicates that GRBs have a standard energy reservoir,
with the apparent variation of $E_0$ reflecting the range of jet
opening angles.

\begin{figure}
\vspace*{0.0mm} 
\includegraphics[width=8.3cm]{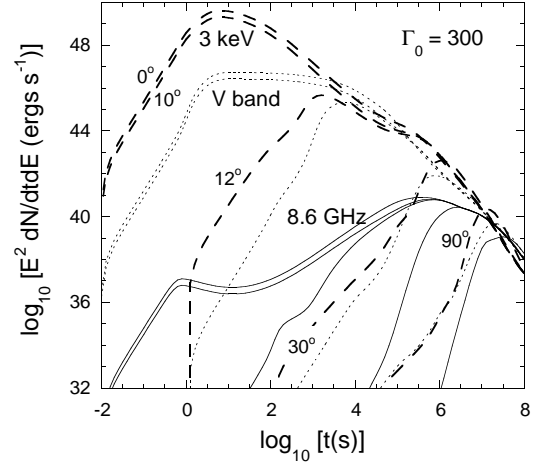} 
\vskip-1.in
\caption{Light curves calculated at various observing energies and
inclination angles $\theta$ for a fireball with parameters the same as
in Fig.\
\ref{fred} and jet opening half-angle $\theta_j = 10^\circ$. The
dashed, dotted, and solid curves give light curves measured at 3 keV,
V band, and 8.6 GHz radio frequencies, and the observing angles
$\theta = 0^\circ, 10^\circ, 12^\circ, 30^\circ$, and $90^\circ$.}
\label{jet}
\end{figure}

\subsection{Beaming in GRBs}

An observer will receive most emission from those portions of a GRB
blast wave that are within an angle $\sim 1/\Gamma$ to the direction
to the observer.  As the blast wave decelerates by sweeping up
material from the CBM, a break in the light curve will occur when the
jet opening half-angle $\theta_j$ becomes smaller than $
1/\Gamma$. This is due to a change from a spherical blast wave
geometry, given by equation (\ref{fe}), to a geometry defined by a
localized emission region, as given by equation (\ref{feblob}).
Assuming that the blast wave decelerates adiabatically in a uniform
surrounding medium, the condition $\theta_j \cong 1/\Gamma =
\Gamma_0^{-1}(x_{br}/x_d)^{3/2} = \Gamma_0^{-1}(t_{br}/t_d)^{3/8}$
implies
\begin{equation}
t_{br} \approx 45(1+z)\;({E_{52}\over n_0})^{1/3}\theta_j^{8/3}\;{\rm days}\;,
\label{tbr}
\end{equation}
from which the jet angle 
\begin{equation}
\theta_j \approx 0.1 [{t_{br}({\rm d})\over 1+z}]^{3/8}\;
({n_0\over E_{52}})^{1/8}
\label{thetaj}
\end{equation}
can be derived \citep{sph99}. Note that the beaming angle is only
weakly dependent on $n_0$ and $E_0$.

Fig.\ \ref{jet} shows calculations of a GRB blast wave in a beamed
geometry \citep{dcm00a} with $\theta_j = 10^\circ$ at different
observing angles and energies. The jet emission is assumed to be
uniform across its surface and not to spread laterally (see
\citet{rho99} when this is not the case). For the parameters used here
($E_{52} = 100$, $e_B = 10^{-4}$, $e_e =0.5$, $n_0 = 100$ cm$^{-3}$,
$p=2.5$), the beaming break occurs at $\sim 10^5$ s, but is obscured
at X-ray and optical frequencies by the appearance of an SSC
component. The appearance of clearly defined breaks in the light
curves of some GRBs, if due to beaming rather than to a variation in
the density of the surrounding medium, limits the ratio $e_e/e_B$ that
defines the strength of the SSC component \citep{dcm00a}.  So-called
``orphan" afterglows, which are those GRBs with jets pointed away from
our line-of-sight which become visible upon deceleration of the blast
wave, may be confused with dirty fireballs \citep{dcm00a,hdl02}.

Afterglow modeling of multiwavelength spectra
  of GRB 980703, GRB 990123, GRB
9905010, and GRB 991216 by \citet{pk01} shows that the CBM is more
consistent with uniform surroundings than with a wind profile with
density $n(r)\propto r^{-2}$. They derive magnetic field parameters in
the range $10^{-4} \lesssim e_B \lesssim 0.05$, electron
energy-transfer parameters in the range $0.01 \lesssim e_e \lesssim
0.1$, and obtain low density surroundings ($\sim 10^{-4}\lesssim
n_0\lesssim 10$ cm$^{-3}$) in their fits. The jet angles $\theta_j$
range from $1^\circ$ to $4^\circ$. The beaming factor for a one-sided
jet is $\approx 1.3\times 10^4/(\theta^\circ)^2$, so that many
misaligned GRBs sources should exist for every detected GRB.

\citet{fra01} have considered a larger sample of GRBs and show that
the beaming results provide evidence for a standard energy reservoir.
Letting $\eta_\gamma$ represent the efficiency for transforming total
energy $E_{tot}$ (which should be distinguished from the apparent
isotropic energy release $E_0$) to $\gamma$ ray energy, then $E_{tot}
\simeq \theta_j^2 E_{\gamma}$(iso)/(4$\eta_{\gamma})$, where
$E_{\gamma}$(iso) is the apparent isotropic energy release in the form
of GRBs. They find that the mean energy release by the GRB sources is
$\langle E_{GRB}\rangle \cong 3\times 10^{51} /(\eta_{\gamma}/0.2)$
ergs, and that the actual event rate of GRBs is $\approx 500 \times$
the observed rate.  If correct, this result has important implications
for the nature of the sources of GRBs.

\section{Source Models}

Keeping in mind that only members of the class of long-duration GRBs
have measured redshifts, considerable evidence connects GRBs to
star-forming regions \citep{djo01,sok01} and, consequently, to a
massive star origin. For example, the associated host galaxies have
blue colors, consistent with galaxy types that are undergoing active
star formation. GRB counterparts are found within the optical radii
and central regions of the host galaxies \citep{blo99a}. Lack of
optical counterparts in some GRBs could be due to extreme reddening
from large quantities of gas and dust in the host galaxy
\citep{gw01}. Supernova-like emissions have been detected in the
late-time optical decay curves of a few GRBs \citep{rei99,blo99b}, and
X-ray evidence for Fe K$\alpha$-line signatures indicates that large
quantities of highly enriched material are near the sources of
GRBs. The spatial and temporal coincidence of GRB 980425 with SN
1998bw, a Type Ic supernova, if true, also connects the sources of
GRBs to SNe \citep{kul98,pia99}.

\subsection{Coalescing Compact Objects}

A central attraction of this model is that binary neutron stars, such
as the Hulse-Taylor pulsar PSR 1913+16, exist. This system has a
merger time scale $\approx 3\times 10^8$ yrs, which is much less than
the Hubble time, so that these events undoubtedly occur. The timescale
for the final coalescence event is very short, with the bulk of the
energy released over a few tens of milliseconds \citep{jan99}.  Such
events should, however, occur in both spiral and elliptical galaxies,
whereas the host galaxies of known GRBs generally exhibit active star
formation and are therefore associated with spiral or disturbed galaxy
hosts.  The absence of counterparts far from the disks of galaxy hosts
also conflicts with this model, as such old stellar systems can travel
great distances before coalescence.  In particular, they should leave
dusty regions where young and massive stars are born.

Merger-rate estimates of neutron-star/neutron-star in addition to
neutron-star/black-hole binaries range from $\approx
10^{-5}$-$10^{-6}$ yr$^{-1}$ in our Galaxy, but are quite uncertain
due to the sensitive dependence of merging timescale on initial binary
separation \citep{nps91,ty93}. Even so, if GRBs outflows are highly
beamed, event rates $\gtrsim 10^{-4}$ yr$^{-1}$ are implied (see \S
6.3). Taken together, these various lines of evidence do not support a
model of coalescing compact objects for the long duration GRBs.

\subsection{Hypernova/Collapsar Model}

In the original formulation of this model \citep{woo93}, a collapsar
was a ``failed" supernova in the sense that a core-collapse event
failed to collapse to form a neutron star and instead produced a black
hole. In order to produce a long-duration GRB with complex pulse
structure, \citet{woo93} argued that accretion had to proceed over a
period of time comparable to the prompt phase of a GRB. The progenitor
star was suggested to be a rotating Wolf-Rayet star that produces,
upon collapse, an accretion disk of several tenths of Solar
masses. Because of the accretion geometry, jetted emission from
neutrino annihilation was formed along the axis of the system.

\citet{pac98} introduced the word ``hypernova" to refer to the
extremely luminous prompt GRB and afterglow events, and concurred
with Woosley that GRBs are formed by the collapse of a massive
star to a black hole surrounded by a massive disk and torus.  He also
suggested that the precursor X-ray emission observed with Ginga in
several GRBs \citep{mur91} was a signature of a dirty fireball
produced as the ejecta cleared away overlying baryonic material, after
which a highly relativistic, clean fireball could then emerge. In the
version of the hypernova model advanced by \citet{pac98}, the events
are in fact as energetic as $10^{54}$ ergs and $10^4$-$10^5$ times rarer
than SNe.

Hydrodynamical simulations of collapsars have specifically treated,
for example, the evolution of a 35 $M_\odot$ main-sequence star whose
14 $M_\odot$ helium core collapses to form a 2-3 $M_\odot$ black hole
\citep{mw99}.  Provided that the core has a large amount of angular
momentum, a delayed accretion event can be formed by the infalling
matter. The black-hole jet simulation is performed by injecting $\sim
10^{51}$ ergs s$^{-1}$, which is argued to be generated by
magneto-hydrodynamical processes rather than neutrino annihilation,
insofar as the latter process produces inadequate energy to break
through the core. In cases where breakthrough is not possible, a
choked event is formed, which could in principle be detected from
neutrino observations 
\citep{mw01}. A distinction is made between prompt and delayed
collapse to a black hole, with black-hole production occurring in the
latter case due to delayed fallback onto a young neutron star
\citep{mwh01}.

The collapsar model must contend with the difficulty of ejecting
baryon-clean material through an overlying shell of material. This is
accomplished through an active central engine that persists for at
least as long as the prompt phase of the GRB. Formation of
relativistic jets of baryonic-clean material reaching $\Gamma_0\sim
10^2$-$10^3$ represents a major difficulty in these models
\citep{tan01}.

\subsection{Supranova Model}

A central motivation of the supranova model of \citet{vs98} is to
identify a site that is originally free of baryon contamination. This
occurs through a two-step collapse to a black hole, where a
``supramassive" neutron star (i.e., with mass exceeding the
Chandrasekhar limit) is formed in the first-step through a supernova
explosion. The neutron star is initially stabilized against collapse
by rotation. The loss of angular momentum support through magnetic
dipole and gravitational radiation leads to collapse to a black hole
after some months to years. A two-step collapse process means that the
neutron star is surrounded by a supernova shell of enriched material
which can explain rebrightening events, as seen in GRB 970508
\citep{vie99}. Alternately, the neutron star could be driven to
collapse by accreting matter in a binary system \citep{vs99}.

The period of activity of a highly magnetized neutron star preceding
its collapse to a black hole can produce a pulsar wind bubble
consisting of low density, highly magnetized pairs \citep{kg01}, in
accord with afterglow model fits \citep{pk01}. The earlier supernova
could yield $\sim 1 M_\odot$ of Fe in the surrounding vicinity. The
discovery of variable Fe absorption in GRB 990705 during the prompt
emission phase \citep{ama00} and X-ray emission features in the
afterglow spectra of GRB 991216 \citep{pir00} provide a test of this
model. Variable X-ray absorption is essentially undetectable from a
GRB surrounded by a uniform CBM, even if highly enriched
\citep{bot99}. Extremely clumped ejecta
\citep{bfd02} or blueshifted resonance scattering 
in a high-velocity outflow \citep{laz01} can account for the variable
absorption.

Fig.\ 9 illustrates the allowed parameter space in the supranova
model, where $f$ refers to the volume filling factor of the ejecta.
By contrast, collapsar progenitors must produce $\gtrsim 0.1 M_\odot$
of iron within $\sim 1$ pc, but massive stars that have metal rich
winds lose too much mass to form a collapsar \citep{fry01}. A
He-merger model, where a binary compact object merges with the He core
of a companion, requires an unreasonably short merger timescale
\citep{bfd02}.

\subsection{Other Models}

These three classes of models hardly exhaust the possibilities. Some
alternative suggestions include a strong-field millisecond pulsar
origin, where the collapse of a white dwarf to a neutron star produces
an $\sim 10^{15}$ Gauss magnetar with a millisecond period
\citep{uso92,uso94}.  Others include the ``cannonball" model
\citep{dd00} involving highly relativistic and highly beamed jets of
materials from SNe. Another involves the transition of a neutron star
in a low-mass binary X-ray system to a strange star \citep{cd96}.
In others, a pair electromagnetic pulse \citep{ruf01}
or the Blandford-Znajek effect \citep{lbw00} extract the energy during the 
black-hole formation event.

\begin{figure}[t]
\vspace*{0.0mm} 
\includegraphics[width=8.3cm]{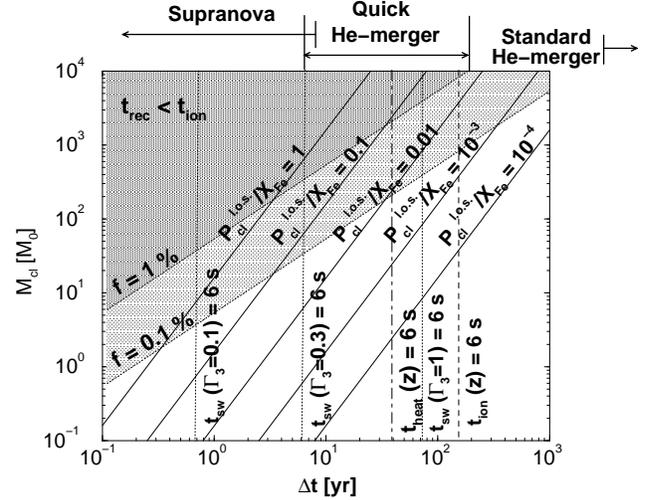} 
\caption{Parameter space giving the amount of 
supernova ejecta mass $M_{cl}$ that must be concentrated in clumps
versus the time delay $\Delta t$ between the primary supernova
explosion and the subsequent GRB to explain the variable absorption
feature in GRB 990705 \citep{bfd02}. Solid lines give the probability
$P_{cl}^{los}$ that a clump will be along the line of sight, which
depends on the iron enhancement $X_{Fe}$ with respect to Solar system
values.  The vertical dotted lines refer to GRB blast wave Lorentz
factors $\Gamma_0 = 1000\Gamma_3$ that give results consistent with
the observations.}
\label{Iron}
\end{figure}

\subsection{Best-Bet Model}

In light of the Fe line observations and the constant energy reservoir
result of \citet{fra01}, the supranova model seems to offer certain
advantages over a collapsar model to explain the origin of GRBs. For
example,
\begin{enumerate}
\item The narrow distribution of energy releases follows 
from a progenitor model involving neutron
stars that collapse to black holes, since the neutron-star progenitor
masses must be very similar.  By contrast, a large diversity of core
masses is possible in the collapsar model, and the varying mantle
structure would strongly modulate the energy radiated as $\gamma$
rays.
\item The immediate environment is much cleaner in the vicinity of a
collapsing neutron star than in a massive-star collapsar model. 
Thus the baryon-contamination problem is more easily solved 
in the supranova model.
\item The production of large amounts of iron in 
the GRB environment is possible in a two-step collapse 
process, with the delay between the first- and second-step event
permitting the nickel and cobalt to decay to iron prior to the GRB.
\item The lack of evidence for winds from massive stars and the low
 surrounding densities is contrary to expectations
in a model where GRBs originate in the direct collapse of massive
stars to black holes \citep{pk01}. Moreover, the low densities
inferred for the surroundings are much less than expected for typical
ISM environment, and could be expected in pulsar wind bubbles
\citep{kg01}.
\item The typical $E_{pk}$ energies observed in
 BATSE can be explained in a prompt external shock 
model, which is compatible with the prompt collapse of a neutron star
to a black hole in the supranova model. The narrow distribution of
$E_{pk}$ energies seems to require fine tuning in a model with
internal shocks in a relativistic wind, and no argument from first
principles has been advanced to explain the typical durations of the
long-duration GRBs in a collapsar wind model.
\end{enumerate}

The evidence is not yet compelling to rule out either model; for
example, iron line observations can be explained in the context of the
collapsar model through a persistent source of emission involving
either a magnetar or the delayed accretion onto a black hole
\citep{rm00,mr01}. By contrast, the delay between neutron star
formation and collapse to a black hole in the supranova model seems to
require ad hoc assumptions about the spin rates and magnetic fields of
the neutron stars formed in SNe explosions. Various anisotropic and clumpy
ejecta are required. If GRB history is any guide, the best bet is on
a model yet to be proposed.

\section{Gamma Ray Burst Cosmology}

We mention some of the main lines of research in GRB cosmology, making
no claims for completeness.  The rate of lensing events on GRBs can
set limits to the cosmological abundance of dark matter in the form of
compact objects, depending on the specific cosmology \citep{mar99}.
High-redshift GRBs provide probes of the early universe that can be
used to examine the redshift dependence of metallicity, the onset of
the first generation of stars, the evolution of Ly$\alpha$ forest and
metal absorption line systems, and the epoch of reionization
\citep{lr00,cl00}. An interesting feature of the afterglow is that its
brightness is not strongly dependent on redshift. The metallicity also
may depend on the location of the GRB source in the host galaxy
\citep{rlb02}. X-ray obscuration of high-redshift GRBs and the
fraction of optically dark bursts will reveal whether GRBs are hosted
in  luminous infrared galaxies similar to the high-redshift SCUBA
galaxies \citep{ram02a}.

\citet{tot97,tot99} and \citet{wij98} 
showed that if GRB sources have a massive star origin, then
the redshift distribution of GRBs yields the star formation rate (SFR)
history of the universe. Gamma rays are less attenuated during
transport than optical radiation, so that a complete sample of GRB
redshifts will arguably give a better measure of the SFR than
measurements inferred from optical and UV surveys of galaxies, which
are subject to strong extinction corrections.

Before a large number of GRB redshifts were known, it was first
necessary to determine whether the BATSE count rate distribution was
in accord with the assumption that the rate of GRBs followed the SFR
history, as inferred, for example, from HST observations
\citep{mpd98}. This depends sensitively on the assumed luminosity
function of GRBs. \citet{kth98} demonstrated, however, that a variety
of luminosity and redshift distributions of cosmological GRBs are
consistent with the observed peak flux distribution, so that a GRB
rate density distribution that followed the SFR of the universe was
not unique. The degeneracy of the unknown GRB luminosity function can
be broken by assuming a specific model for GRB emission, as done by
\citet{bd00} for an external shock model. This leads to a specific
prediction for the redshift distribution of GRBs (Fig.\ 2, lower
panel) based upon the assumed SFR function, and the GRB energy and
Lorentz factor distributions. Comparison with GRB redshift data can
then be used to refine the model parameters and test the basic
assumption that the rate density of GRBs follows the SFR of the
universe.

GRBs can have important effects on the galaxy environment, and were
proposed as the energy sources of HI shells and stellar arcs
\citep{eeh98,lp98}, though the evidence for beaming makes these
possibilities less likely. GRBs could melt dust grains by GRB UV
radiation to produce flash-heated chondrules in the early Solar system
\citep{mh99}. In one sense, the most cosmologically interesting effect
of GRBs is their influence on biological activity \citep{sw02} and
life extinction \citep{dd01}, which depends not only on GRB photon
production but on particles accelerated by GRBs, namely cosmic rays.

\section{Cosmic Ray Production by GRBs}

As summarized in \S 4, several lines of evidence indicate that GRBs
are closely related to a subset of SNe that drives a relativistic
outflow in addition to the nonrelativistic ejecta expelled during the
collapse of the massive core to a neutron star. The relativistic
ejecta decelerate to nonrelativistic speeds at the Sedov radius by
sweeping up matter from the external medium.  Particle acceleration
occurs at these shocks, just as in the nonrelativistic shocks from
``normal" Type Ia and Type II supernova remnants, though with the
addition of a relativistic phase of deceleration. The first-order
shock Fermi mechanism is generally recognized as the mechanism that
accelerates GeV/nuc cosmic rays in the converging flows formed by the
forward shock. In addition, second-order Fermi acceleration of
particles through gyroresonant particle-wave interactions with
magnetic turbulence generated in the shocked fluid can also accelerate
particles to high energies.

We consider the simplest case of particle acceleration via the
first- and second-order Fermi processes at the external shock formed
by ejecta decelerating in a uniform surrounding medium. These
considerations will establish whether GRBs can accelerate particle to
the ankle of the cosmic ray spectrum at $3\times 10^{18}$-$10^{19}$ eV
and to ultra-high ($>10^{19}$ eV) energies. We show that ultra-high
energy cosmic rays (UHECRs) can in principle be accelerated by GRB
blast waves. Additional effects involving internal shocks,
acceleration at the reverse shock, or a highly magnetized CBM can
improve the acceleration efficiency.

\subsection{Constraints on Acceleration}

A basic requirement for particle acceleration to some maximum energy
$E_{max}$ in the observer frame is that the Larmor radius
$r^\prime_{\rm L}$ be smaller than the comoving size scale $r^\prime$
of the system. The comoving Larmor radius $r^\prime_{\rm L} =
mc^2\gamma/qB$ for particles with mass $m = Am_p$ and charge $q =
ze$. The requirement that $r^\prime_{\rm L} \leq r^\prime$ implies
that
\begin{equation}
E_{max,{\rm L}} = {qB\Gamma\delta c t_{var}\over 1+z} \cong
10^{18}\;{ZB({\rm G})\Gamma_{300}^2 t_{var}({\rm s})\over 1+z}\; {\rm
eV}\;,\;\label{Emax}
\end{equation}
where $r^\prime$ is related to the measured variability timescale
$t_{var}$ through the relation $t_{var} \cong
(1+z)t^\prime_{var}/\delta \cong (1+z)r^\prime/c\delta$, and $\delta
\approx \Gamma$. Recalling from equation (\ref{B^2}) that $B \cong 116
\sqrt{e_Bn_0}\Gamma_{300}$ G, we see that UHECR acceleration can be
achieved for highly relativistic GRB outflows, especially if large $Z$
nuclei such as iron are accelerated.

In Fermi processes, particles can increase their energy by $\sim mc^2
\gamma$ on a Larmor timescale $t_{\rm L} = 2\pi r_{\rm L}/c = 2\pi
mc\gamma/qB$ \citep{rm98}.  Hence the maximum acceleration rate is
$\dot \gamma_{acc} = f_{acc}(qB/2\pi mc)$, where $f_{acc} \lesssim
1$. A basic constraint on particle acceleration is the radiation
reaction limit due to synchrotron losses. The synchrotron loss rate
\begin{equation}
-\dot\gamma_{syn} = {4\over 3} c \sigma_{\rm T}\;({B^2 \over 8\pi m_e
c^2})\;{Z^4\over (m/m_e)^3}\;\gamma^2\;.
\label{gammasyn}
\end{equation}
Equating this with $\dot\gamma_{acc}$ gives the radiation-reaction
limit on particle acceleration through Fermi processes
\begin{equation}
E_{max,syn}= mc^2 \Gamma\;\sqrt{{3ef_{acc}\over \sigma_{\rm T}
B}}\;{(m/m_e)\over Z^{3/2}}
\label{Emaxsyn}
\end{equation}
This gives $E_{max,syn}\approx 6\times
10^{15}\Gamma_{300}\sqrt{f_{acc}/B({\rm G})}$ eV for electrons,
and a maximum
synchrotron photon energy $h\nu_{max} $ $\cong$ $ 20 f_{acc}\Gamma$ MeV,
independent of $B$, as is well-known \citep{jag96}. For ions, the
synchrotron radiation-reaction limiting energy is
\begin{equation}
E_{max,syn~ion}= 2.4\times 10^{22} {A^2\over
Z^{3/2}}\Gamma_{300}\sqrt{f_{acc}\over B({\rm G})}\;{\rm eV}
\label{Emaxsynion}
\end{equation}
The maximum ion synchrotron photon energy is $h\nu_{max} \cong 40
(Af_{acc}/Z^2)\Gamma$ GeV. Other processes, such as Compton losses for
leptons and photomeson and photopair losses for ions, can limit
acceleration further. Provided that these losses are not severe, the
synchrotron radiation reaction is not a severe limitation for UHECR
particle acceleration.

Another kinematic limitation on first-order Fermi acceleration arises
from the size scale of the system. Because particles must be scattered
upstream of the shock, the upstream Larmor radius $r_{\rm L}$ must be
smaller than the characteristic size of the system during the period
when the shock is strong. The Sedov radius, given by equation
(\ref{l_S}), is the radius where the blast wave has swept up an amount
of matter equal to the ejecta mass. Particle acceleration is strongly
suppressed after the explosion enters the Sedov regime. The condition
$r_{\rm L} < \ell_{\rm S}$ implies that
\begin{equation}
E_{max,size} \approx 2\times 10^{15} Z\;B_{\mu{\rm G}}\;({m_\odot\over
n_0})^{1/3}\;{\rm eV}\;
\label{Emaxsize}
\end{equation}
where the $B_{\mu{\rm G}}$ and $n_0$ are the mean magnetic field,
in $\mu$G, and density of the interstellar medium.

The time required to cycle across the supernova shock provides a
further limitation on particle acceleration by shock Fermi processes.
The first-order Fermi acceleration rate $\dot \gamma_{FI}$ $ \approx $ $ 
\beta\gamma/t_{cyc}$, where $\beta$ is the shock speed, and
the cycle time 
\begin{equation}
t_{cyc} \cong {4\over c} ({\kappa_-\over u_-}+{\kappa_+\over
u_+})\; \approx {\eta r_{{\rm L},-}\over \beta c}\;.
\label{tcyc}
\end{equation}
Here $\kappa$ is the diffusion coefficient, $u$ is the speed of the
flow in the rest frame of the shock, and subscripts $-$ and $+$ refer
to quantities upstream and downstream of the shock, respectively
\citep{dru83,kir94}. The expression on the right-hand-side assumes
that the time to scatter is determined by quantities upstream of the
shock, and that the diffusion coefficient is given by the Bohm
diffusion limit $\eta r_{\rm L,-}$. Differences in cycle times from
quasi-parallel and quasi-perpendicular shocks arise from shock-drift
in the latter case. The maximum energy-gain rates per unit
distance $x$ for nonrelativistic quasi-parallel and
quasi-perpendicular shocks, respectively, are given by
\begin{equation}
{d\Ep\over dx}|_{\parallel,max} \simeq \beta q B_-\;\;;\;\;
\label{dEpdxa}
{d\Ep\over dx}|_{\perp,max} \simeq \beta^{2/3} q B_-\;\;.\;\;
\label{dEpdxb}
\end{equation}
The energy increases by only
a factor $\cong 2$ in a single
cycle following the first cycle in
relativistic shock acceleration \citep{gal99}. 

By integrating the energy-gain rate over the equation 
of blast wave evolution given by equation (\ref{P(x)}), 
the classic result of \citet{lc83} is generalized to give
the maximum particle energy 
\begin{equation}
E_{max,I}\approx 10^{16} Z B_{\mu{\rm
G}}\beta_0^{2/3}({m_\odot\Gamma_0\over n_0})^{1/3}\;{\rm eV}
\label{Emaxt}
\end{equation}
that can be obtained by first-order Fermi acceleration at a forward
external shock. This shows that neither nonrelativistic or
relativistic first-order shock-Fermi processes are capable of
accelerating particles to the ankle of the cosmic ray spectrum.  In
fact, nonrelativistic ($\beta_0\Gamma_0 \ll 1$) shocks have
difficulties to accelerate particles past the knee of the spectrum
\citep{lc83}, although winds from the pre-supernova star can enhance
the local magnetic field and reduce the surrounding medium density and
so permit acceleration to much higher energies \citep{vb88}.

Second-order Fermi acceleration from gyroresonant interactions between
particles and waves in relativistic shocks can accelerate particles to
considerably higher energies than first-order Fermi acceleration when
the wave speeds approach $c$. For a wave spectrum composed of Alfven
waves with speed $v_{\rm A}$, the acceleration rate from a power law
spectrum of waves $w(k) \propto k^{-v}$ with wavenumber $k>k_{min}$ is
given by
\begin{equation}
\dot\gamma_{FII} \approx ({v_{\rm A}\over c})^2 \xi 
(v-1)\;({c\over r_{\rm L}^0})(r_{\rm L}^0 k_{min} \gamma)^{v-1}\;.
\label{dotgammaFII}
\end{equation}
Here $r_{\rm L}^0 = mc^2/qB$ is the nonrelativistic Larmor radius, and
$v$ is the index of the turbulence spectrum ($v = 5/3$ corresponds to
a Kolmogorov spectrum). Magnetic field turbulence is produced in the
relativistic shocked fluid when particles and dust are captured and
isotropized \citep{ps00,sd00}. After integrating over the energy gain
rate during the duration of a decelerating blast wave, one obtains a
maximum particle energy
\begin{equation}
E_{max,II} \approx 8\times 
10^{20} Z K_v e_B^{1/2}n_0^{1/6} f_\Delta \beta_0 
(m_\odot \Gamma_0)^{1/3}\;{\rm eV}\;
\label{EmaxII}
\end{equation}
\citep{der01}, where $K_v \cong [0.6 e_B 
\beta_0\xi /f_\Delta]^{1/(2-v)}$ for $v = 3/2$,$5/3$, and
$f_\Delta x/\Gamma^2\cong x/12\Gamma^2$ gives the observer frame width
of the blast wave from the shock-jump conditions in a uniform CBM
\citep{vie98,vie95,pmr98,dh01}.  The strong dependence of the maximum
energy on $\beta_0$ makes second-order processes unimportant in
nonrelativistic flows when $\beta_0 \ll 0.1$.

\begin{figure}[t]
\vskip-1.2in
\vspace*{0.0mm} 
\includegraphics[width=8.3cm]{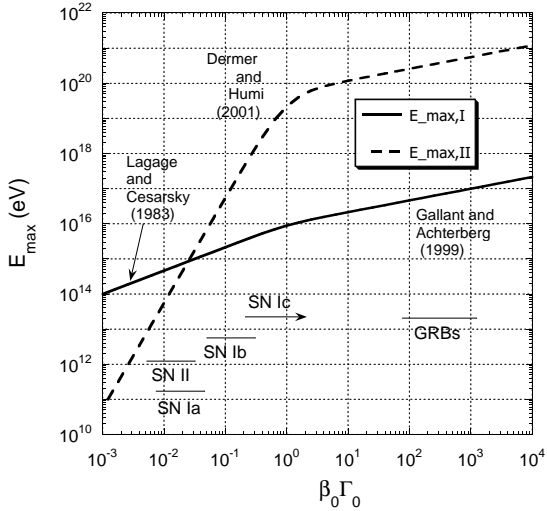} 
\vskip-0.3in
\caption{Maximum proton energies from first- and second-order Fermi 
acceleration for explosions with initial Lorentz factors $\Gamma_0$
and speed $\beta_0 c$. }
\label{Emaxfig}
\end{figure}

Fig.\ \ref{Emaxfig} compares the maximum proton energies that can be
obtained by first-and second-order Fermi acceleration at an external
shock formed by an explosion with total energy $m_\odot = 1$,
corresponding to $1.8\times 10^{54}$ ergs.  The surrounding external
medium contains a 1 $\mu$G field and a density of 1 cm$^{-3}$.  A
fully turbulent magnetic field with $\xi = 1$, $e_B = 0.2$,
 and $v=3/2$ is assumed
in the second-order process.  First-order Fermi processes cannot
accelerate particles to ultra-high energies for particles accelerated
from low energies for typical ISM conditions.  Relativistic shocks can, however, give a
pre-existing relativistic particle population a single-cycle boost by
a factor $\Gamma^2$ \citep{gal99,vie95}.

\subsection{UHECR by GRBs}

The Larmor radius of a particle with energy $10^{20}E_{20}$ eV is
$\approx 100 E_{20}/(ZB_{\mu{\rm G}})$ kpc. Unless UHECRs are heavy
nuclei, it is unlikely that they originate from our Galaxy.  The
proposal that UHECRs are accelerated by extragalactic GRB sources was
first advanced by \citet{vie95}, \citet{wax95}, and \citet{mu95}. The
first two of these authors noted that the energy density of UHECRs was
comparable to the energy density that would be produced by GRB sources
within the GZK radius, assuming that the $\gamma$-ray and UHECR power
from GRBs are roughly equal. Although the details of this estimate are
incorrect inasmuch as they were made before the Beppo-SAX results on
the detailed redshift distribution of GRBs, the overall energy-density
estimate still gives strong circumstantial support for this
hypothesis.

The argument proceeds by noting that the energy density of UHECRs
observed near Earth is simply
\begin{equation}
u_{UH} \cong \zeta\; {L_{GRB}t_{esc}\over V_{prod}}\;
\label{u_UH}
\end{equation}
where $L_{GRB}$ is the power of GRBs throughout the production
volume 
$V_{prod}$ of the universe. UHECRs are produced with 
an efficiency $\zeta$ compared with the $\gamma$-ray power
and ``escape" from the universe
primarily through photohadronic processes with an effective
escape time $t_{esc}$.

\begin{figure}[t]
\vskip-0.7in
\vspace*{0.0mm} 
\includegraphics[width=8.3cm]{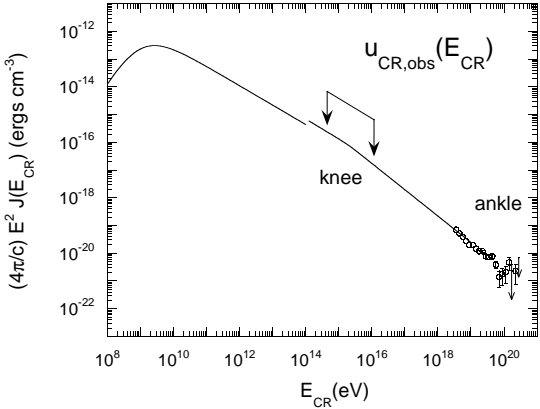} 
\vskip-1.in
\caption{Cosmic ray energy density observed near Earth. The two solid
 curves show extrapolations to model fits of the observed cosmic-ray
 proton \citep{sim83} and all-particle \citep{fow00} spectrum, and the
 data points are UHECR observations from AGASA \citep{tak98}.  }
\label{CR}
\end{figure}

The mean $\gamma$-ray fluence of BATSE GRBs is $F_\gamma \approx
3\times 10^{-6}$ ergs cm$^{-3}$ and their rate over the full sky is
$\dot N_{GRB}\approx 2/$day. If most GRBs are at redshift $\langle z
\rangle \sim 1$, then their mean distance is $\langle d \rangle
\approx 2\times 10^{28}$ cm, so that the average isotropic energy
release of a typical GRB source is $\langle E_\gamma \rangle \approx
4\pi \langle d \rangle ^2F_\gamma/(1+z) \cong 8\times 10^{51}$ ergs,
implying a mean GRB power into the universe of $L_{GRB} \approx
2\times 10^{47}$ ergs s$^{-1}$. (This estimate is independent of the
beaming fraction, because a smaller beaming fraction implies a
proportionately larger number of sources.)  UHECR protons lose energy
due to photomeson processes with CMB photons in the reaction $p+\gamma
\rightarrow p+\pi^0, n+\pi^+$. The effective distance for $10^{20}$ eV
protons to lose 50\% of their energy is 140 Mpc \citep{sta00}, so that
$t_{esc} \cong 1.5\times 10^{16}$. This implies that the energy
density observed locally is $u_{UH} \approx 10^{-22} \eta$ ergs
cm$^{-3}$. By comparing with Fig.\ \ref{CR}, we see that this estimate
is about an order of magnitude below the energy density of super-GZK
particles with energies exceeding $10^{20}$ eV.

Given the limitations of this crude estimate, the coincidence between
the GRB power and the UHECR energy density suggests that this
hypothesis may be correct if GRBs can accelerate UHECRs with high
efficiency ($\eta \approx 1$).  An improved estimate must consider
several other aspects, including
\begin{enumerate}
\item The efficiency of GRB sources to produce $\gamma$-ray emission within
the BATSE band. If only 10\% of the total energy of GRBs are radiated
at soft $\gamma$-ray energies, then these sources will be much more
energetic on average, which will improve the comparison -- provided
that UHECR acceleration remains very efficient.
\item The actual redshift distribution of GRB sources.
If the rate density of GRB sources
follows the SFR history of the universe \citep{mpd98}, then the local
comoving rate density will be $\approx 6$-10 times less than the rate
density at $z \approx 1$. This effect, when factored into our improved
knowledge of the GRB redshift distribution showing that the faintest
GRBs are at $z >1$, makes the hypothesis that UHECRs are accelerated
by GRBs more difficult to sustain \citep{ste00}.
\item The existence of the dirty and clean fireball 
classes of GRBs. These sources classes would not trigger BATSE, and
the dirty fireball class could be detected as the X-ray rich GRBs
\citep{hei01}. This correction could increase the average power into
the universe by a factor $\sim 3-5$ \citep{der00}.
\item The uncertainty in the energy density of the super-GZK cosmic 
rays. The UHECR energy
density shown in Fig.\ \ref{CR} is from the AGASA
measurements. Monocular HiRes observations show a lower flux than
AGASA above $\approx 10^{20}$ eV by a factor of $\approx 5$
\citep{som02}, which improves the coincidence between the UHECR energy
density and the GRB power.
\end{enumerate}
A detailed estimate of GRB power is given by \citet{der00} in the
context of an external shock model for GRBs performed by
\citet{bd00}. The conclusion of this study is that GRBs remain a
viable source origin for UHECRs. A testable prediction of the
hypothesis is that star-forming galaxies which host GRB activity will
be surrounded by neutron-decay halos \citep{der00}, as discussed in \S 7.

\subsection{Rate and Power of GRBs in the Milky Way}

The preceding estimates were made for extragalactic sources of cosmic
rays. GRBs will also take place in galaxies such as our own. After a
relativistic GRB blast wave decelerates to nonrelativistic speeds, it
evolves in a manner similar to nonrelativistic supernova remnant
shocks. GRBs will therefore power cosmic rays with energies below the
knee of the cosmic ray spectrum. These cosmic rays will add to cosmic
rays accelerated in the remnants of Type I and Type II SNe. Here we
estimate the rate and power of GRBs throughout the Milky Way in order
to determine the contribution of GRBs to cosmic ray production.

We assume the constant energy reservoir result of \citet{fra01} to
estimate the rate of GRBs into $L^*$ galaxies such as the Milky
Way. The local density of $L^*$ galaxies can be derived from the
Schechter luminosity function, and is $\approx 1/$(200-500 Mpc$^3$)
\citep{wij98,sw02,der00}. The BATSE observations imply, as already
noted, $\sim 2$ GRBs/day over the full sky. Due to beaming, this rate
is increased by a factor of $500 f_{500}$, where $f_{500} \sim
1$. Given that the volume of the universe is $\sim 4\pi(4000$
Mpc)$^3/3$, this implies a rate per $L^*$ galaxy of $$
\rho_{L^*}\approx {300~{\rm Mpc}^3/L^*\over {4\pi\over 3}
(4000~{\rm Mpc})^3}\;{2\over {\rm day}}\;{365\over {\rm yr}}
\times 500 f_{500}\times SFR\times K_{FT}$$
\begin{equation}
\;\;\;\;\;\;\;\approx 2\times 10^{-4}\;({SFR\over 1/6})
\times ({ K_{FT}\over 3})\times f_{500} {\rm ~yr}^{-1}\;.
\label{rhoL*}
\end{equation}
The factor $SFR$ corrects for the star-formation activity at the
present epoch. The SFR rate per comoving volume is about 1/6 as active
now as it was at $z = 1$, and the factor $K_{FT}$ accounts for dirty
and clean fireball transients that are not detected as GRBs. Thus a
GRB occurs about once every 5000 years throughout the Milky Way.

\begin{table}
  \begin{center}
    \caption{Ejecta Speeds and Rates of Supernovae and
GRBs in the Milky Way}\vspace{1em}
    \renewcommand{\arraystretch}{1.2}
    \begin{tabular}[h]{lccc}
      \hline
     Explosion  &  Outflow & $\langle \beta_0\Gamma_0\rangle$ & Rate  \\
   Type    &  Speed (km s$^{-1}$) &  & (century$^{-1}$)  \\
      \hline
      SN Ia & $\lesssim 2\times 10^{4}$ & 0.03 & 0.42   \\
      SN II & $\sim 10^3$-$2\times 10^{4}$ & 0.01 & 1.7   \\
      SN Ib/c & $\sim 1.5\times 10^3$-$2\times 10^{5}$ & 0.2 & 0.28   \\
      Dirty & $3\times 10^{5}$ & 30 & ?   \\
      GRB & $3\times 10^{5}$ & 300 & $\sim 0.02$   \\
      Clean & $3\times 10^{5}$ & 3000 & ?   \\
      \hline \\
      \end{tabular}
    \label{tab:table1}
  \end{center}
\end{table}

Table 1 gives the outflow speeds and rates of different types of SNe
and GRBs in the Galaxy. The data for the outflow speeds of Types Ia,
II, and Ib/c come from \citet{loz92} and \citet{wei00}.  The mean
values of the initial dimensionless momentum $\langle \beta_0\Gamma_0
\rangle$ of the outflows are also given. The supernova rate data are
obtained by multiplying the results of \citet{cap99} for galaxies of
Type Sbc-Sd (renormalized by \citet{pan00} to an H-band luminosity for
$h_{70} = 1$) by a factor of 2 to account for the luminosity of the
Milky Way. One sees that the GRB rate is about 10\% as frequent as
Type Ib/c SNe, and about 1\% as frequent as Type II SNe. The rates of
dirty and clean fireballs are unknown, but if the X-ray rich
$\gamma$-ray bursts comprise the tail of the dirty fireball
population, then they could be even more frequent than the GRB
population. Statistical fits \citep{bd00} to the BATSE data suggest
that the rate of clean fireballs is smaller than the GRB rate.

By weighting the GRB rate, equation (\ref{rhoL*}), by the mean energy
of $3\times 10^{51}/(\eta_\gamma/0.2)$ ergs, we see that the
time-averaged power of GRBs throughout the Milky Way or other $L_*$
galaxies is
\begin{equation}
L_{MW} \approx 2\times 10^{40}  {f_{500}\;\over 
(\eta_\gamma/ 0.2)}\; 
({SFR\over 1/6})\;({ K_{FT}\over 3})\;{\rm ergs}\;{\rm s}^{-1}\;.\;
\label{LMW}
\end{equation}
Because each GRB has a total energy release of a few $\times 10^{51}$
ergs, comparable to SNe, but occur $\sim 100$ times less frequently,
the available power from the sources of GRBs is only a few per cent of
the power from Type Ia and Type II SNe.  This would suggest that GRB
sources can only make a minor contribution to cosmic ray production
below the ankle, but the efficiency for accelerating cosmic rays in
the nonrelativistic SN outflows could be considerably less than in the
relativistic outflows of GRBs.

In this regard, note that Cas A has recently been detected with HEGRA
at TeV energies at the level of $\approx 8\times 10^{-13}$ ergs
cm$^{-2}$ s$^{-1}$ \citep{aha01}. Spectral considerations suggest that
the emission is hadronic. The expected level of 100 MeV- TeV
$\gamma$-ray production can be estimated to be $$F_\gamma \approx
(4\pi d^2)^{-1}\times 10^{51}{\rm ~ergs~}\eta_p\times c\sigma_{pp}
n_0\approx$$
\begin{equation}
7\times 10^{-11}\;({\eta_p\over 0.1})\; 
n_0\;({d\over 3.4 {\rm ~kpc}})^{-2}{\rm ~ergs~cm}^{-2}{\rm ~s}^{-1}\;,\;
\label{Fgamma}
\end{equation}
where the strong interaction cross section $\sigma_{pp} \approx 30$
mb. This expression may underestimate the spectral energy flux
measured at a fixed frequency range and the $\gamma$-ray production
efficiency, but unless the density is unusually tenuous in the
vicinity of Cas A and other SNRs examined with the Whipple telescope
\citep{buc98}, the TeV observations suggest that most SNRs do not
accelerate cosmic rays with high efficiency.

The hypothesis that cosmic rays are accelerated by SNRs is favored in
view of the power provided by SNe \citep{gai90}. With an energy
release of $\sim 10^{51}$ ergs per SN and a rate of $\approx 1$ every
30 years, the power into the Galaxy from SNe is at the level of
$\approx 10^{42}$ ergs s$^{-1}$, well in excess of the $\approx
5\times 10^{40}$ ergs s$^{-1}$ required to power GeV/nuc cosmic
rays. Moreover, first-order Fermi acceleration by a strong shock
produces a $-2$ spectrum, similar to the injection spectrum required
to explain the measured cosmic ray spectrum prior to steepening due to
propagation effects.

In spite of this apparent success, observational difficulties for the
hypothesis that cosmic rays are accelerated by SNe bear repeating
\citep{der00}
\begin{itemize}
\item The unidentified EGRET $\gamma$-ray sources
have not yet been firmly associated with 
SNRs and do not display distinct $\pi^0$ features \citep{esp96};
\item The spectrum of the diffuse galactic 
$\gamma$-ray background is harder than is
expected if the locally observed cosmic ray proton and ion spectrum is
similar throughout the Galaxy \citep{hun97};
\item The acceleration of particles above the 
knee of the cosmic ray spectrum is difficult to 
explain in the simplest theory \citep{lc83} of cosmic 
ray acceleration by SNRs; and
\item As we have seen, TeV radiation is not detected at the expected levels.
\end{itemize}

In view of these difficulties, it seems possible that cosmic rays are
predominantly accelerated by the subset of SNe which eject
relativistic outflows. Bright enhancements of hadronic emission from
those SNe which host GRB events are implied. From the rate estimates
shown in Table 1, we see that about 1 in 20 to 1 in 100 SNRs would
exhibit this enhanced emission.  The better imaging and sensitivity of
the {\it GLAST} telescope and the next generation of ground-based air
Cherenkov telescopes will be able to test this hypothesis.

\section{High Energy Neutrons and Neutrinos from GRBs}

Observations of neutral particles and their secondary 
radiations will test the hypothesis that
cosmic rays are accelerated by GRB blast waves. Neutron-decay halos are
formed around galaxies that harbor GRB activity
\citep{der00}. Equation (\ref{LMW}) shows that the sources of GRBs
inject a mean power of $10^{40\pm 1}$ ergs s$^{-1}$ into a typical
$L^*$ galaxy. In order to explain the locally observed UHECR energy
density, each GRB source must accelerate ions with high
efficiency. Neutrons will be directly produced through the process
$p+\gamma \rightarrow n+\pi^+$ in the GRB blast wave and will escape
the acceleration site.  The neutrons, unbound by the magnetic field in
the blast wave, leave the acceleration site with Lorentz factors
$\gamma_n = 10^{10}\gamma_{10}$, where $0.1 \lesssim \gamma_{10}
\lesssim 100$.  The neutrons decay on a timescale $\gamma_n t_n \cong
3\times 10^5\gamma_{10}$ yrs, where the neutron $\beta$-decay lifetime
$t_n \cong 900$ s.

The neutrons travel a characteristic distance $\lambda_n \cong 90
\gamma_{10}$ kpc before they decay into highly relativistic electrons
and protons.  The electrons initiate an electromagnetic cascade that
forms a diffuse $\approx 10$ TeV Compton $\gamma$-ray component and a
nonthermal synchrotron component extending to soft $\gamma$-ray
energies.  The halo is unfortunately very weak. Three orders of
magnitude reduction in the injection luminosity results from The
$\beta$-decay to an electron results in three orders of magnitude
reduction of power, and a further reduction results from the
efficiency of photomeson production in the GRB blast wave. In an
external shock model, the efficiency can be lower than 1\%, implying
powers of neutron decay radiation halos surrounding $L^*$ galaxies at
 $\sim 10^{35}$ ergs s$^{-1}$.

Will water and ice Cherenkov neutrino telescopes detect GRBs in the
light of high-energy photohadronic neutrinos?  The comoving photon
energy density is
\begin{equation}
\ep u^\prime(\ep ) \cong {d_L^2\over 2cr_b^{\prime~2}}\; 
{f_\e \over \Gamma^2}\;,\;{\rm and}\;\e \cong {2\Gamma \ep\over 1+z}\;
\label{epkupk}
\end{equation}
for a blast-wave geometry. The proper-frame density increases by $4\Gamma$
upon crossing a relativistic shock. The invariance of
the total number of particles implies that
$N =x^2 (\Delta\Omega)\Delta^\prime_{sh} n^\prime$ $=$ $(\Delta\Omega)
n_0\int_{x_0}^x d\bar x \bar x^2$, so that
\begin{equation}
r_b^\prime \cong \Delta_{sh}^\prime = {x^3-x_0^3\over 12\Gamma x^2} \rightarrow
\cases{{x\over 12\Gamma}  ,& uniform CBM \cr\cr
        {x-x_0\over 4\Gamma} , & ${x-x_0\over x_0 }\ll 1 $ \cr}.
\label{rbprime}
\end{equation}
 The comoving photon density is $n^\prime(\ep ) = \ep u^\prime(\ep
)/m_e c^2 \e^{\prime~2}$, and the synchrotron emission is assumed to
be isotropic in the comoving frame, so that $n^\prime(\ep , \mu^\prime
)\cong {1\over 2} n^\prime (\ep )$.

The time scale for significant energy loss by photohadronic reactions is
\begin{equation}
t^{\prime~-1}_{p\gamma\rightarrow \pi}\simeq {c\over 5}
\int_0^\infty d\ep \int_{-1}^1 d\mu^\prime 
(1-\mu^\prime ) n^\prime (\ep ,\mu^\prime)
\sigma_{p\gamma\rightarrow \pi}(\e^{\prime\prime})\; , \;
\label{nprime}
\end{equation}
where $\sigma_{p\gamma\rightarrow \pi}(\e^{\prime\prime})\cong
\sigma_0\delta[\mu^\prime - (1-\e_\Delta/\gamma_p^\prime\ep )]$,
$\sigma_0\approx 2\times 10^{-28}$ cm$^2$, $\epsilon_\Delta \approx
640$, $\e^{\prime\prime} = \gamma_p^\prime\ep (1-\mu^\prime )$, and we
now use primes to refer particle Lorentz factors to their proper
frame. The comoving time available to undergo hadronic reactions is
$t_{ava}\cong r_b^\prime/c \cong 2\Gamma t/(1+z)$. The quantity
\begin{equation}
\eta \equiv {t^\prime_{ava}\over t^\prime_{p\gamma\rightarrow \pi}}
\cong {\sigma_0d_L^2\e_\Delta\over 20 \gamma_p^\prime m_ec^3\Gamma^2
r_b^\prime}\;\int_{\epsilon_\Delta/2\gamma_p^\prime}^\infty
d\ep\;\e^{\prime-3}\; f_\e\;
\label{eta}
\end{equation}
represents an efficiency for converting accelerated particle energy
into neutrinos. The threshold condition $\gamma_p^\prime \ep \cong
\e_\Delta$ requires acceleration of protons with observer-frame
energies $\gamma_p \cong \Gamma^2
\epsilon_\Delta/[(1+z)\e_{pk}]\propto \Gamma^{-2}\psim t^{3/4}$ from
scattering with photons at the $\nu F_\nu$ peak photon energy
$\e_{pk}$. As can be seen, demands on particle acceleration are more
easily satisfied during the prompt phase of the GRB. Neutrinos are
formed in the observer's frame with energies of $\approx 0.05 
m_pc^2\gamma_p$. The peak photon energy of the $\nu F_\nu$
spectrum is
\begin{equation}
\e_{pk} \cong {2\Gamma\over (1+z)}\e_B\gamma_{min}^2\cong
{500\over 1+z}\;k_p^2 e_e^2 (e_Bn_0)^{1/2}\Gamma_{300}^4\;.
\label{epknew}
\end{equation}
The brightest neutrino production episode occurs $t\approx 
t_d$ for a fast cooling scenario, as can be shown from
equation (\ref{eta}). Quantities evaluated at $t_d$ wear
hats. Equations (\ref{fe}) and (\ref{Ne})
or rule (\ref{Pi0}) \citep{dcb99} 
imply
\begin{eqnarray}
\nonumber
\hat f_{\e_{pk}} \simeq {2\Gamma^2\over 4\pi d_L^2}\;
({B^2\over 8\pi}c\sigma_{\rm T})\;N_e^0\gamma_c\gamma_{min}
\simeq {3\over 2}\,{k_p e_eE_0\over 4\pi d_L^2t_d}\\
~~~~~~~~\simeq 10^{-6}\;
{E_{52}^{2/3}\Gamma_{300}^{8/3}n_0^{1/3}k_p e_e\over d_{28}^2 (1+z)}\;
{{\rm ergs}\over{\rm ~cm}^{2}{\rm s}}\;,
\label{epk17}
\end{eqnarray}
so $\hat f_\e = \hat f_{\e_{pk}}(\e/\hat \e_{pk})^{\alpha_\nu}$ and
 $\alpha_\nu = 1/2$ for $\e\leq \hat \e_{pk}$, $\alpha_\nu = (2-p)/2$
 for $\e > \hat \e_{pk}$ in the fast cooling limit.  From eq.\ (\ref{eta}),
\begin{equation}
\hat\eta  \simeq {\sigma_0d_L^2\e_\Delta \hat f_{\e_{pk}}\over 5 
\gamma^\prime_{pk} m_ec^3(1+z)^2 \hat \e^2_{pk} r_b^\prime}\;
\int_{\hat y_{pk}}^\infty dy\; y^{\alpha_\nu - 3}\;.
\label{etahat}
\end{equation}
At $\hat y_{pk} = 1$, where 
$\alpha_\nu = \alpha_{\nu,u}=(2-p)/2$, 
\begin{eqnarray}
\nonumber
\hat\eta_{fc} \simeq {\sigma_0 d_L^2 \hat 
f_{\e_{pk}}\over 5(2-\alpha_{\nu,u})f_\Delta x_d m_ec^3 (1+z)\hat\e_{pk}}\\
~~~~~~~~~\simeq {10^{-2}\over (2-\alpha_{\nu,u})(1+z) 
f_\Delta}\;{n_0^{1/6}E_{52}^{1/3} \Gamma_{300}^{-2/3}
\over k_p e_e e_B^{1/2}}\;,\;
\label{etapk}
\end{eqnarray}
letting $r_b^\prime = f_\Delta x_d/\Gamma$.

The low efficiencies of neutrino production makes detection of GRBs
with a 1 km$^{3}$ neutrino telescope extremely unlikely within the
context of a {\it uniform} external shock scenario \citep{der00}. Low
Lorentz-factor dirty fireballs have increased  efficiencies $
\hat\eta_{fc}$ by as much as three orders of magnitude, 
but require acceleration to $\gamma_{pk} \simeq
10^{10}/\Gamma_0^{2}k_p^2 e_e^2\sqrt{e_B n_0}$, which follows from the
relation $\hat y_{pk} = \Gamma_0^2 \e_\Delta/\gamma_{pk}(1+z)\hat
\e_{pk}=1$. The fast-cooling condition $\gamma_c < \gamma_{min}$
implies $e_e e_B\gtrsim $ $4\times 10^{-5}
/[E_{52}^{1/3}n_0^{2/3}k_p(1+z) \Gamma_{300}^{4/3}]$, which
additionally shrinks available parameter space.

Result (\ref{etapk}) gives the photomeson production efficiency at the
deceleration time scale for a proton with energy
$m_pc^2\gamma_{pk}$ $\simeq$ $ 10^{14}$ eV$/[k_p^2 e_e^2 \sqrt{e_B
n_0}\;\Gamma_{300}^2$] eV in a fast cooling blast wave. Secondary
neutrinos with energy $\sim 15$\% $\times m_pc^2 \gamma_{pk}$ would be
received from such GRBs. Deep water (Baikal, Nestor, Nemo, Antares)
and ice (AMANDA, IceCube) neutrino detectors have detection
probability $P\sim 10^{-4}P_{-4}$, with $P_{-4} \sim 1$ at PeV
neutrino energies. The number of detected neutrinos $N_\nu$ can be
approximated by the product of the telescope area $10^{10} A_{10} $
cm$^{2}$, the neutrino flux $\cong 10^{52}E_{52}$ $/[4\pi d_L^2
(m_pc^2\gamma_{pk})]$ cm$^{-2}$ s$^{-1}$, the multiplicity $3/2$ for
muon neutrinos, and a bandwidth correction factor $0.1\zeta_{-1}$.
The threshold energy for photomeson production by protons interacting
with photons at the peak of the $\nu F_\nu$ synchrotron spectrum at $t
= t_d$ is given by $m_pc^2\gamma_{pk} \cong m_pc^2
\Gamma_0^2\e_\Delta/[(1+z)\hat\e_{pk}]$ $\cong 170$ ergs$/[k_p^2
e_e^2 \sqrt{e_Bn_0}\;\Gamma_{300}^2]$.

Multiplying these factors together gives an estimate of the number of
neutrinos that would be detected from a GRB, assuming that a comparable
amount of energy is in the nonthermal proton spectrum as in the
explosion:
\begin{equation}
N_\nu\simeq 4\times 10^{-4} \;{A_{10} P_{-4}E_{52}^{4/3}
 k_p e_e n_0^{2/3}\Gamma_{300}^{4/3} \zeta_{0.1}
\over
d_{28}^2 (2-\alpha_{\nu,u})(1+z)f_\Delta}\;.
\label{Nnu}
\end{equation}
Equation (\ref{Nnu}) is consistent with the numerical calculations of 
\citet{der00}, but much less accurate by neglecting the energy dependence of 
$P$, and neutrino production in the afterglow phase. This result shows
 that there is no prospect for detecting $\gtrsim 1 $ neutrinos from
 FRED-type GRBs, and we verify the computations \citep{der00} of
 the weak neutrino flux produced by GRB blast waves decelerating in a
 uniform CBM.

Equation (\ref{Nnu}) says that GRBs are much brighter $\nu$
sources when $\Gamma_{300} \gg 1$, $n_0 \gg 1$, and $f_\Delta \ll
1$. Clean fireballs are however rare \citep{bd00}. Density
inhomogeneities from the SN envelope in the supranova model could
enhance $\nu$ production in an external shock model, not only through
the density effect indicated in equation (\ref{Nnu}), but through
photo-hadronic processes with 
external photons \citep{ad01} and back-scattered photons. Eq.\
(\ref{rbprime}) shows that thin shells are formed in a clumpy medium,
so that $f_\Delta$ could be $\ll 1/12$.\footnote{This answers a
criticism of the external shock model posed by R.\ Sari.} Because
spiky and highly variable GRBs are due to large density contrasts in
the external shock model on a size scale $\ll x/\Gamma$ \citep{dm99},
it seems possible that the spikiest and most variable GRBs could
produce detectable high-energy $\nu$ emission in an external shock
model.  A definitive statement must await further studies.

Internal shock-model advocates \citep{wax95,wb97}
determine the blast-wave width from the observed variability time
scale $t_{var} \simeq (1+z)r_b^\prime/(c\delta)$, so that $r_b^\prime
= f_\Delta x/\Gamma \cong 2c\Gamma t_{var}/(1+z)$. Thus $f_\Delta
\simeq t_{var}/t_d$, that is, 
the shell must lie within a distance $ct_{var}\ll ct_d$ of the central 
source, where the deceleration time $t_d$ is associated with any surrounding
CBM.  Sufficiently small shell widths,
though larger than the size scale of the original explosion, can enhance
$\nu$ production considerably, so that the observable $\gamma$-ray
power sets the only available scale on energy release.  Pair
attenuation in these inner regions will make $\gamma$-$\gamma$
spectral cutoffs in the GLAST energy range. Studies of
$\gamma$-$\gamma$ opacity during the prompt phase of an external shock
model find transparency up to TeV energies \citep{dcm00a}.

In the uniform external shock model, as we have seen, a very low
efficiency for neutrino production is a consequence of the low
radiation energy density in the comoving frame \citep{der00}. In
models involving internal shocks, by contrast, the neutrino production
efficiency can be much higher due to the larger radiation density
\citep{wb97,wb00}. In the latter case, GRBs should be detectable with
km$^3$ scale neutrino detectors if UHECRs are indeed
accelerated by GRB blast waves. Photon emission from the reverse shock
can also enhance neutrino emission \citep{dl01}. Neutron decoupling
from protons in the expanding fireball can also produce a strong flux
of 5-10 GeV neutrinos \citep{bm00}. Secondary production can also be effective
for high-energy neutrino production \citep{sch01}.

We think that no measurable neutrino flux will be detected from
smooth-profile GRBs.  Spiky GRBs could be neutrino-bright, that is,
potentially detectable in a km$^3$-scale experiment.  It will be
important to have continuous sky coverage with a GRB/hard X-ray survey
instrument as IceCube and the northern hemisphere
water arrays accumulate data.

\section{Advanced Blast Wave Theory}

All aspects of GRB theory cannot be treated in a short review, but at
least some mention should be made of the main subjects overlooked. The
physics of colliding shells in an internal shock/wind scenario was
omitted, in keeping with Ockam's razor: ``{\it pluralitas non est
ponenda sine neccesitate,}" that is, ``entities should not be
multiplied unnecessarily."  This hypothesis is not needed in GRB
research; for an opposite viewpoint, see
\citet{piran99}, and for a balanced assessment, see
\citet{mes02}. (Colliding shell physics is of course important
in black-hole jet studies; see, e.g.,  \citet{spa01}.)

The physical approach summarized here has been developed by dozens of
authors, and represents a generalization of supernova remnant physics
to explosions with relativistic ejecta. Some important early
contributions leading to the relativistic synchrotron-shock model
include the works of \citet{kat94} and \citet{tav96}.  A departure
from a nonthermal synchrotron origin for the prompt emission is made
by \citet{laz00} and \citet{ghi00}, who argue that it is instead due
to bulk Compton upscattering in a high photon energy-density region in
the funnel of an exploding star.  Part of the motivation for their
work was observations of very hard X-ray emission with $\nu F_\nu$
indices $\alpha_\nu \gtrsim 2$ during the prompt $\gamma$-ray luminous
phase of 5-10\% of GRBs \citep{cri97,pre98}.  This strongly
contradicts the optically-thin synchrotron shock model, which predicts
that only spectra with $\alpha_\nu \leq 4/3$ can be radiated in a
nonthermal synchrotron model.

Elaborations on the 
synchrotron-shock theory that may account for such hard spectra include
photospheric emission \citep{mr00} and radiation reprocessing by a
medium heated by the $\gamma$ rays \citep{db00}. These latter effects
could lead to sites of enhanced annihilation radiation in the ISM at
the location of GRB remnants, which INTEGRAL could
discover. \citet{mt00} and \citet{tm00} treated
Compton-backscatter effects from evolving $\gamma$-ray photon fronts,
though initially approximated as an $\sim$10-100 lt-s shell
of photons. The Compton
back-scattered photons provide targets for successive waves of
incident GRB photons through $\gamma\gamma$ pair-production
interactions.  \citet{bel01} and \citet{mrr01} consider how
pair-loading and preacceleration (``surfing") of the medium ahead of
the $\gamma$-ray light front can alter the early afterglow and produce
a bright optical/UV emission component during the early phases of a
GRB. These early optical flashes might have been mistaken for a reverse
shock emission.
\citet{pac01} suggests that these effects could produce optical flashes 
preceding a GRB. 

Finally, we note that the theory of the central engine and
gravitational radiation from GRBs is not treated here.

\section{Conclusions}

The Introduction summarizes the scope of this review, and will not be
repeated here.  The themes that I have tried to stress, and that would
not have been possible to include in a review written only a year ago,
are the implications of afterglow studies and X-ray absorption and
line observations on source models.  In particular, the constant
energy reservoir result of \citet{fra01}, derived by modeling breaks
in GRB afterglow light curves, implies that GRBs are much more
numerous and much less energetic than earlier believed. GRB sources
are probably related to an unusual type of supernova occuring at
$\sim 1$\% of the rate of Type II SNe, with energy releases several
times greater than Type Ia or Type II SNe.  Relativistic shock waves
can accelerate particles to the ankle of the cosmic ray spectrum
through second-order Fermi processes, so that cosmic rays between the
knee and ankle of the cosmic ray spectrum could originate from GRB
sources.  If relativistic shocks accelerate particles with
much greater efficiency than nonrelativistic shocks, then GRB sources
could make a significant contribution to the production of cosmic rays
with energies from GeV/nuc to the knee.

Cosmic rays may originate from the subset of supernovae that eject
relativistic outflows. Gamma-ray observations with GLAST and
ground-based air Cherenkov observatories will be fundamentally
important to identify the sources of cosmic rays by searching for
hadronic emission from discrete sites in the Galaxy. The Swift
Observatory, in conjunction with ground-based followup, will establish
a large database from which GRB redshifts, energies, and beaming
fractions will be obtained. Neutrino observatories may force GRB
source models to be revised. We look forward to further advances that
will make this review obsolete.

\begin{acknowledgements}
I thank E. Ramirez-Ruiz, N. Trentham, D. Lazzati, M. J. Rees,
P. M\'esz\'aros, M. Vietri, R. Schlickeiser, E. Waxman, and especially
M. B\"ottcher for conversations and e-mail correspondence in
preparation for this talk and during the completion of this
review. The usual disclaimer applies: all errors are my own. 

This work
is supported by the Office of Naval Research and the NASA Astrophysics
Theory Program (DPR S-13756G).
\end{acknowledgements}

\end{document}